\documentclass[11pt]{article}

\usepackage[]{acl}

\usepackage{times}
\usepackage{latexsym}

\usepackage[T1]{fontenc}
\usepackage[utf8]{inputenc}

\usepackage{microtype}

\usepackage{inconsolata}

\usepackage{graphicx}

\usepackage{booktabs}
\usepackage{makecell}

\usepackage{multirow}

\usepackage{algorithm}
\usepackage{algorithmic}

\usepackage{booktabs}
\usepackage{array}
\usepackage{longtable}
\usepackage{amsmath}
\usepackage[most]{tcolorbox}
\usepackage{xcolor}
\usepackage{pifont}

\newcommand{\nop}[1]{}

\title{HARP: Hierarchical Adaptive Ranking with Preference-Adaptive Fusion for Query-Based CVE Prioritization}

\author{
Haochen Liu\thanks{Work done during an internship at NEC Labs America.}$^{1}$,
Zhengzhang Chen\thanks{Corresponding author.}$^{2}$,
Haoyu Wang$^{2}$,
Yanchi Liu$^{2}$,
Jundong Li$^{1}$,
Haifeng Chen$^{2}$ \\
$^{1}$University of Virginia \\
$^{2}$NEC Laboratories America \\
\texttt{sat2pv@virginia.edu},
\texttt{zchen@nec-labs.com},
\texttt{haoyu@nec-labs.com}, \\
\texttt{yanchi@nec-labs.com},
\texttt{jundong@virginia.edu},
\texttt{Haifeng@nec-labs.com}
}

\usepackage{xcolor}
\usepackage{amssymb}
\usepackage{amsfonts}

\usepackage[most]{tcolorbox}
\usepackage{xcolor}
\usepackage{amssymb}
\usepackage{tabularx}
\usepackage{array}

\begin{document}
\maketitle

\begin{abstract}
Vulnerability prioritization is inherently preference dependent, since the same CVE can receive different remediation priority under different operational preference scenarios. Existing scoring systems and ranking methods typically assume a fixed criterion. In practice, organizations already operate under a preference scenario, but this preference is often implicit and difficult to express as a written prompt instruction, while triage queries usually do not encode it. Past validated triage cases under the current scenario are more readily available. We study query-based CVE prioritization in this setting and propose HARP, a graph-grounded multi-view framework that ranks candidates from a natural-language query together with a support bank of historical labeled examples from the current preference scenario, without requiring an explicit textual summary of that scenario. HARP retrieves evidence from a vulnerability knowledge graph, scores candidates with policy-conditioned global, enterprise, and user views, and fits view-fusion weights from sampled supports. Experiments across three preference scenarios and multiple backbone LLMs show that HARP outperforms multiple baselines, expressing our method's effectiveness.

\end{abstract}

\section{Introduction}

Modern software systems continuously accumulate vulnerabilities across large software ecosystems. With limited remediation resources, organizations must decide which CVEs (Common Vulnerabilities and Exposures) to remediate first. This prioritization process depends on multiple factors, including exploitability, affected products, deployment environment, patch availability, and user-facing exposure~\cite{li2017patches,desmale2023firehose}.

Existing vulnerability information such as CVSS, EPSS, CISA KEV, and SSVC~\cite{cvss31,ssvc,epss2023} provide useful signals about severity, exploitation likelihood, and remediation urgency, but they do not define a single universal prioritization strategy. In practice, ranking depends on the organization's current preference scenario. The same CVEs may therefore require different rankings under different scenarios, even when the triage query looks similar.%, shown in Figure~\ref{fig:intro}.

\begin{figure*}[t]
    \centering
    \includegraphics[width=1\linewidth]{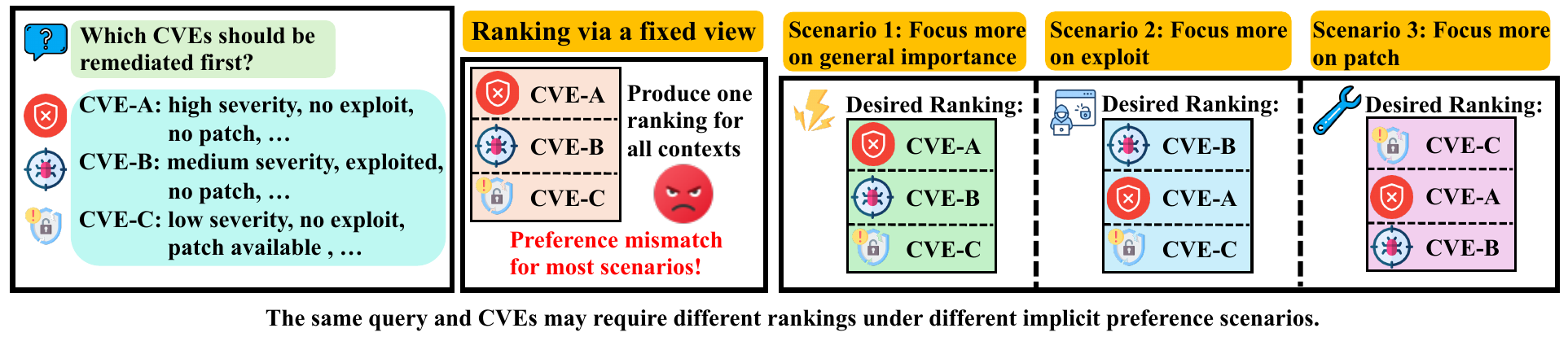}
    \caption{Motivating example of preference-dependent CVE prioritization. The same query and candidate CVEs can require different rankings under different preference scenarios. Fixed scoring or single-prompt ranking produces one ordering and may fail to match the current scenario's remediation preference.}
    \label{fig:intro}
\vspace{-15pt}
\end{figure*}

Existing vulnerability prioritization methods typically rely on fixed industrial signals~\cite{allodi2014risk,sabottke2015vulnerability,bozorgi2010beyond,suciu2022expected,wudi2022diffcvss}, lexical matching or embedding-based retrieval ranking~\cite{burges2005ranknet,burges2010lambdamart,karpukhin2020dpr,khattab2020colbert}, or LLM prompting with optional retrieved evidence~\cite{lewis2020rag}. However, these approaches generally assume a fixed or explicitly specified ranking criterion, making them poorly suited when remediation priority depends on the real preference scenario.

Recent parameter-efficient adaptation methods show that different model behaviors can be represented through lightweight specialized modules without full retraining~\cite{houlsby2019adapter,hu2022lora}. Related work on knowledge editing likewise studies localized parameter updates for modular behavior control~\cite{meng2022rome,meng2023memit,wang2024knowledgeediting,wu2024reft,liu2026representation,zhang2026loka,zhang2026mind}. Although our focus is preference adaptation rather than factual knowledge editing, these lines of work motivate using policy-specific lightweight modules to capture different prioritization preferences while keeping a shared backbone frozen.

The central challenge is that preference scenarios are often implicit and complex. Although an organization already operates under a remediation objective, that objective is difficult to express as a written preference description for prompting or as a clean supervision label for training. Queries likewise usually omit such a description. As a result, methods that require an explicit textual preference scenario are often impractical. In contrast, past validated triage cases under the current scenario are typically available and can serve as concrete evidence of the desired ranking tendency. Moreover, a single fixed scoring view is usually insufficient: remediation decisions jointly depend on complementary evidence such as global severity and exploitability, enterprise exposure and patch feasibility, and different preference scenarios require different balances among these factors. This motivates a multi-view formulation rather than ranking under one fixed criterion. Figure~\ref{fig:intro} illustrates this contrast between ranking with a fixed view and ranking under different preference scenarios.

In this paper, we propose HARP, a \underline{H}ierarchical \underline{A}daptive \underline{R}anking with Preference-Adaptive Fusion framework for query-based CVE \underline{P}rioritization. HARP retrieves candidate CVEs and evidence from a vulnerability knowledge graph, applies policy-conditioned multi-view scoring, and combines view scores through adaptive fusion with Top-$K$ consensus routing. HARP adapts to the current preference scenario through historical support examples, without requiring an explicit textual summary of that scenario. %We construct a query-based CVE ranking benchmark with manually annotated rankings under three preference scenarios, each with its own bank of past labeled examples, as detailed in Section~\ref{sec:evaluation}.
We construct a query-based CVE ranking benchmark with ground-truth rankings annotated by cybersecurity practitioners under three preference scenarios. These evaluation labels are independent of the policy-specific supervision used for adapter training, and each scenario is associated with its own bank of historical labeled examples, as detailed in Section~\ref{sec:evaluation}. Although the policy templates and evaluation scenarios are motivated by common real-world remediation objectives, they were designed independently and serve different purposes: the former provide supervision for adapter training, whereas the latter define qualitative human evaluation criteria.

Our contributions are summarized as follows:
\vspace{-7pt}
\begin{itemize}
    \item We study query-based CVE prioritization under implicit preference scenarios and propose HARP, which adapts ranking with policy-conditioned multi-view scoring and historical support-guided fusion without requiring a written preference description.
    \vspace{-7pt}
    %\item We construct a query-based CVE ranking benchmark with test rankings manually annotated under three preference scenarios, together with manually constructed policy-specific supervision for training.
    \item We construct a query-based CVE ranking benchmark with ground-truth rankings annotated by cybersecurity practitioners under three preference scenarios, together with separately designed policy-specific supervision for adapter training.
  
    \vspace{-7pt}
    \item We conduct experiments showing that HARP consistently improves ranking quality over baselines across multiple backbone LLMs.
\end{itemize}

\section{Problem Formulation}

\nop{We study query-based vulnerability prioritization in a knowledge-graph-grounded setting. Let $G=(V,E)$ denote a vulnerability knowledge graph, where nodes represent CVEs and related entities such as products, vendors, and weaknesses. Given a natural-language query $q$, the system constructs a query-specific candidate set
\begin{equation}
    C(q)=\{c_1,c_2,\ldots,c_n\}\subseteq V,
\end{equation}
where each $c_i$ is a candidate CVE node retrieved from the graph. The goal is to assign each candidate a priority score such that candidates with higher scores receive higher remediation priority.

Vulnerability priority is defined with respect to an implicit decision context. The same CVE may receive different priorities when the underlying preference emphasizes different factors. Since this preference is not directly observed, we model it as a latent context variable $cxt$ and write the ideal scoring function as
\begin{equation}
    F_{cxt}(q,c_i).
\end{equation}
Here, $cxt$ controls how candidates are evaluated and compared, but it cannot be directly observed or used as a supervision signal. The learning problem is therefore to estimate a context-dependent ranking function without explicit access to the underlying decision context.}

We study query-based vulnerability prioritization in a graph-grounded setting. Let $G=(V,E)$ denote a vulnerability knowledge graph, where nodes represent CVEs and related entities such as products, vendors, and weaknesses. Given a natural-language query $q$, the system constructs a query-specific candidate set
\begin{equation}
C(q)=\{c_1,c_2,\ldots,c_n\}\subseteq V,
\end{equation}
where each $c_i$ is a retrieved candidate CVE. The goal is to assign each candidate a score that reflects its remediation priority.

Ranking depends on an existing preference scenario $\gamma$, corresponding to the organization's current remediation objective. The same query and candidates may require different rankings under different $\gamma$. Because $\gamma$ is typically implicit and is not provided as a written preference description in the query, we represent the current scenario by a support bank $B_{\gamma}$ of previously labeled triage examples collected under that scenario and disjoint from the test instances. The ranking function is
\begin{equation}
F(q,c_i;B_{\gamma}),
\end{equation}
where $q$ is used for retrieval, policy selection, and view scoring, and $B_{\gamma}$ provides historical preference supervision for view fusion. The system receives neither a textual preference description nor a scenario identity token, and it does not predict $\gamma$ as an intermediate variable.

\section{Method}

\begin{figure*}[t]
\centering
\includegraphics[width=1\linewidth]{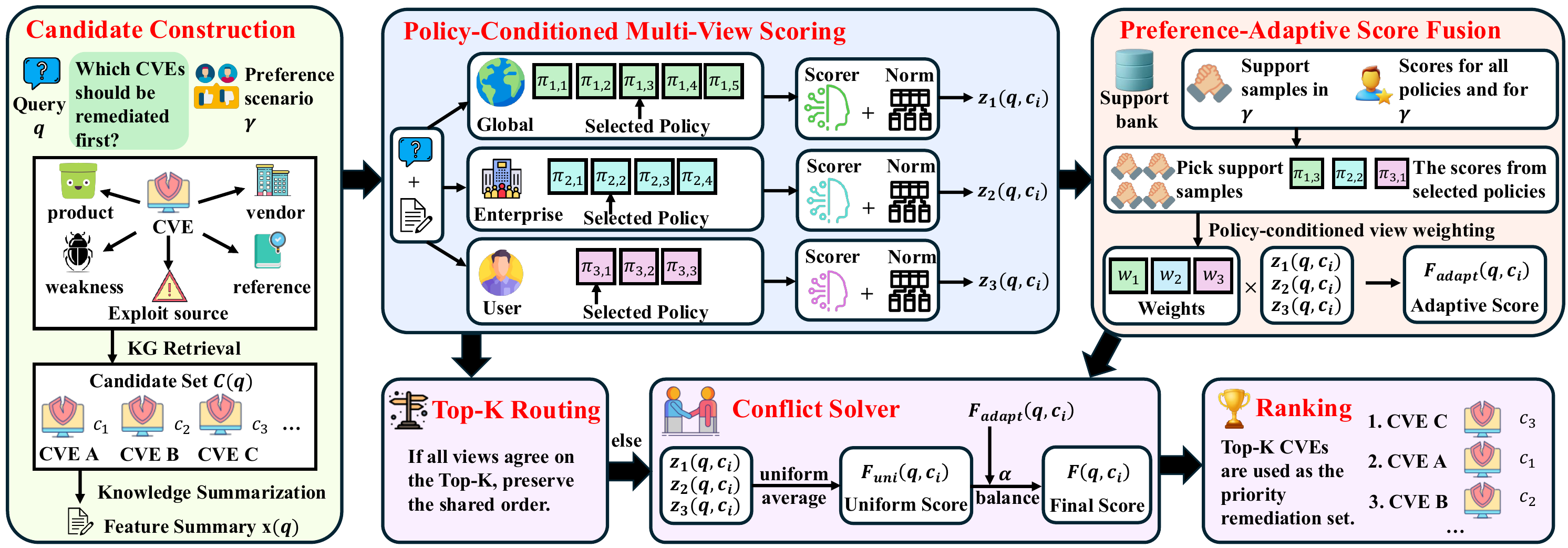}
    \caption{Overview of the HARP framework for query-based CVE prioritization. Given a natural-language query and a support bank of past labeled examples from the current preference scenario, HARP retrieves candidate CVEs and evidence from a vulnerability knowledge graph, scores candidates through policy-conditioned global, enterprise, and user views, adaptively fuses the view-specific scores with support-bank guidance, and produces the final ranked CVE list with multi-view regularization and Top-$K$ consensus routing.}
    \label{fig:pipeline}
    \vspace{-15pt}
\end{figure*}

\nop{To estimate the context-sensitive scoring function defined above, we design a graph-grounded and preference-adaptive prioritization framework. As illustrated in Figure~\ref{fig:pipeline}, HARP consists of four main stages: graph-grounded candidate construction, policy-conditioned multi-view scoring, support-bank-based adaptive fusion, and final multi-view regularized ranking. Given a query $q$, the system retrieves a candidate set from the knowledge graph, scores each candidate from multiple prioritization views under query-selected policies, and adaptively fuses the normalized view-specific scores.

The framework approximates the latent decision context $\gamma$ through two complementary mechanisms. Policy-conditioned scoring adapts the evaluation criterion within each view, while support-bank-based fusion adjusts the contribution of different views using precomputed score patterns from sampled support instances under the selected policy configuration. A uniform multi-view regularization term is further introduced to preserve complementary evidence from all views and prevent the final ranking from collapsing onto a single perspective.}

To estimate the preference-dependent scoring function above, we propose HARP, a graph-grounded and preference-adaptive prioritization framework. As illustrated in Figure~\ref{fig:pipeline}, HARP consists of four main stages: graph-grounded candidate construction, policy-conditioned multi-view scoring, support-bank-based Preference-adaptive fusion, and Top-$K$ consensus routing. Given a query $q$ and a support bank $B_{\gamma}$ of past labeled examples from the current preference scenario, the system retrieves candidates from the knowledge graph, scores them under query-selected policies in each view, and fuses the normalized view scores using sampled supports from $B_{\gamma}$. This design adapts to the current scenario without requiring an explicit textual preference description.

Policy-conditioned scoring adapts the scoring behavior within each view, while support-bank fusion adapts the view mixture using historical ranking supervision from $B_{\gamma}$. %A uniform multi-view regularization term preserves complementary evidence from all views. 
Details of graph construction are in Appendix~\ref{app:kg_construction}, and the support-bank protocol is in Appendix~\ref{app:support_protocol}.

\subsection{Vulnerability Knowledge Graph Construction}

We use an enriched vulnerability knowledge graph as the structured evidence source for retrieval and ranking. The base graph is built from CVE records with nodes for CVEs, products, vendors, weaknesses, and references~\cite{host-etal-2023-constructing}, and is further enriched with exploit- and patch-related external signals. Details of graph construction and enrichment are provided in Appendix~\ref{app:kg_construction}.

\subsection{Candidate Retrieval and Evidence Construction}

Given the enriched graph $G$ and a query $q$, we construct a query-specific candidate set and a shared candidate summary for downstream scoring. The candidate set is retrieved as
\begin{equation}
C(q)=R(q,G),
\end{equation}
where $R$ denotes the graph-based retrieval operator. The query is first aligned with graph entities implied by its content, such as products, vendors, or risk-related terms. Starting from these entities, the system traverses graph relations to collect relevant CVE nodes and forms the candidate set.

We then construct a shared candidate summary for the retrieved candidates. For each candidate in $C(q)$, we collect fields such as description, severity score, exploit signals, affected products, weakness category, vendor information, and references. These fields are organized into a structured input and summarized by an LLM:
\begin{equation}
x(q)=\mathrm{LLM}_{\mathrm{sum}}(C(q)).
\end{equation}

The resulting representation $x(q)$ summarizes the evidence of the candidate set. It is provided to the downstream scoring modules so that each CVE can be evaluated relative to the full candidate set rather than in isolation.

\subsection{Policy-Conditioned Multi-View Scoring}

Given the candidate feature summary $x(q)$, HARP scores each candidate under multiple prioritization views. Because a single scoring rule is rarely sufficient across remediation objectives, we use policy-specific modules to control the scoring behavior within each view.

We consider $L$ prioritization views. In this work, the views correspond to global, enterprise, and user-facing views $(L=3)$. The global view captures general vulnerability risk, such as overall severity and exploitability; the enterprise view focuses on organizational relevance, such as affected products, exposure, and remediation feasibility; and the user-facing view captures end-user impact. This decomposition preserves complementary evidence rather than forcing all prioritization factors into a single scoring criterion.

Within each view $l$, we define a policy set
\begin{equation}
\Pi_l=\{\pi_{l,1},\pi_{l,2},\ldots,\pi_{l,K_l}\},
\end{equation}
where each policy $\pi_{l,k}$ represents a specific evaluation preference under that view. Each policy is implemented as a lightweight adapter module on a shared frozen backbone, rather than as a factual knowledge edit. The number of policies need not be equal across views: different views admit different meaningful prioritization trends. In our benchmark we use five global, four enterprise, and three user policies (Table~\ref{tab:global_policy}); the overall framework remains modular and does not require these counts.

For a query $q$, the model selects one policy for each view. Let $h(q)$ denote the query embedding, and let $\kappa_{l,k}$ be a key vector derived from the textual name of policy $\pi_{l,k}$. The selected policy for view $l$:
\begin{equation}
\pi_l^*=\arg\max_{\pi_{l,k}\in\Pi_l}\mathrm{sim}(h(q),\kappa_{l,k}),
\end{equation}
where $\mathrm{sim}(\cdot,\cdot)$ denotes similarity. This selects the policy whose key is most aligned with the query.

Using the selected policy and the shared candidate feature summary, the view-specific scorer assigns a raw score to each candidate:
\begin{equation}
a_l(q,c_i)=f_l(q,c_i\mid \pi_l^*,x(q)).
\end{equation}

Here, $a_l(q,c_i)$ is the raw score of candidate $c_i$ under view $l$. The output of this stage is a set of raw score vectors
\begin{equation}
\{a_l(q)\}_{l=1}^{L},
\quad
a_l(q)=[a_l(q,c_1),\ldots,a_l(q,c_n)].
\end{equation}

These raw scores represent the ranking evidence produced by each view. In the next stage, they are normalized and adaptively combined to produce the final priority score.

\subsection{Preference-Adaptive Score Fusion}

The view-specific scores produced in the previous stage may have different scales. Directly combining raw scores can make the final ranking dominated by views with larger numerical ranges rather than stronger ranking evidence. We therefore apply per-view $z$-score normalization within each query's candidate set:
\begin{equation}
z_l(q,c_i)=\frac{a_l(q,c_i)-\mu_l(q)}{\sqrt{\sigma_l^{2}(q)+\epsilon}},
\end{equation}
where $\mu_l(q)$ and $\sigma_l^{2}(q)$ are the mean and variance of $\{a_l(q,c_i)\}_{i=1}^{n}$, and $\epsilon$ is a small constant for numerical stability. This makes fusion depend on relative ranking evidence rather than raw score magnitudes.

After normalization, we adaptively combine the view-specific scores. A single global fusion rule would assume that all queries require the same balance among views. In our setting, however, view contributions should follow the current preference scenario, which we represent by a support bank of historical labeled examples. Under the main protocol, each preference scenario $\gamma$ has its own bank $B_{\gamma}$ (three scenarios, three banks). The bank is built from the support data only, where evaluation queries never enter; in deployment it can be refreshed with newly validated triage cases under the same scenario.

For each support query, we have run all policies in all views and store their normalized score vectors together with the ranking supervision:
\begin{equation}
B_{\gamma}=\{(q^{(k)},C^{(k)},z^{(k)},y^{(k)})\}_{k=1}^{M},
\end{equation}
where $q^{(k)}$ is a support query, $C^{(k)}$ is its candidate set, $y^{(k)}$ denotes the ranking supervision associated with that bank, and $z^{(k)}$ stores the normalized score vectors produced by all view-policy scorers.

At inference time, the system attaches $B_{\gamma}$ for the current scenario and uses it only as labeled historical examples: the scoring prompt contains no scenario information, and HARP does not predict one. Let
\begin{equation}
\pi^*(q)=(\pi_{1,r_1^*},\pi_{2,r_2^*},\ldots,\pi_{L,r_L^*})
\end{equation}
denote the policy tuple selected in the previous stage. We then uniformly sample $|S(q)|=3$ supports from $B_{\gamma}$ and, using the stored scores under $\pi^*(q)$, fit
\begin{equation}
w^{\gamma}(q)=
\arg\min_{w}
\sum_{k\in S(q)}
\mathcal{L}\!\left(
y^{(k)},
\sum_{l=1}^{L}w_l z_{l,r_l^*}(q^{(k)})
\right),
\end{equation}
where $\mathcal{L}$ denotes a ranking loss. Because these score vectors are precomputed, sampling three supports and fitting a three-dimensional weight vector is cheap relative to the LLM scoring calls. Policy selection depends on the query, while the preference signal for weighting comes from the historical labels in $B_{\gamma}$. 
The adaptive fused score is
\begin{equation}
F_{\mathrm{adapt}}(q,c_i)=\sum_{l=1}^{L}w_l^{\gamma}(q)z_{l,r_l^*}(q,c_i).
\end{equation}
Although adaptive fusion adjusts view importance, fitted weights may become very small or negative. To preserve complementary evidence from all views, we introduce a uniform multi-view regularization term:
\begin{equation}
F_{\mathrm{uni}}(q,c_i)=\frac{1}{L}\sum_{l=1}^{L}z_{l,r_l^*}(q,c_i),
\end{equation}
\begin{equation}
F(q,c_i)=\alpha F_{\mathrm{adapt}}(q,c_i)+(1-\alpha)F_{\mathrm{uni}}(q,c_i),
\end{equation}
where $\alpha\in[0,1]$ is selected on the development split ($\alpha{=}0.3$ in the main experiments).

After computing the final fused score, HARP ranks candidates by sorting $F(q,c_i)$ in descending order. We additionally apply a Top-$K$ consensus routing rule, with $K=\lceil 0.3\cdot|C(q)|\rceil$. If the selected view-specific rankings produce the same top-$K$ ordering, HARP preserves this shared ordering at the top of the final ranking and orders the remaining candidates by the fused score. Otherwise, HARP ranks all candidates directly by the fused score. This is a lightweight consensus check on top of weighted fusion, rather than a general conflict-resolution procedure.

\subsection{Training}

We train the policy-conditioned scorers using a ranking dataset
\begin{equation}
D=\{(q^{(k)},C(q^{(k)}),u^{(k)})\}_{k=1}^{N},
\end{equation}
where $u^{(k)}$ denotes the feature information of candidates in $C(q^{(k)})$. The backbone LLM is kept fixed, and each policy is trained through its own lightweight adapter. Thus, policies share the same backbone model while learning different scoring behaviors through separate trainable parameters.

For each view $l$ and policy $\pi_{l,r}\in\Pi_l$, we use policy-specific ranking supervision $y_{l,r}^{(k)}$ over the candidate evidence $u^{(k)}$. Different policies emphasize different remediation criteria over the same candidates and therefore induce different training targets. Details of the policy definitions used to guide this construction are provided in Appendix~\ref{app:preference_policy}. %Importantly, these policy targets are used only to train adapters; they are not the rankings used for test evaluation, are not aligned one-to-one with the three preference scenarios, and are never provided to the model as inputs at test time.
Importantly, these policy targets are used only to train the adapters. %Test rankings are instead obtained from independent expert annotations provided by cybersecurity practitioners following scenario-specific prioritization guidelines. The policy templates are never used to construct evaluation labels and are never provided during inference.
Test rankings are instead obtained from independent annotations provided by cybersecurity practitioners following scenario-specific prioritization guidelines. The evaluation scenarios and their annotation guidelines were developed independently of the policy templates used to generate adapter supervision. The policy templates are never used to construct evaluation labels and are never provided during inference.

Adapters therefore learn reusable scoring behaviors rather than a preference-scenario classifier. Preference adaptation for a target scenario is introduced at inference time through that scenario's historical support bank $B_{\gamma}$, while the query is used for retrieval, policy selection, and view scoring.

Let $\theta_{l,r}$ denote the trainable parameters for policy $\pi_{l,r}$ in view $l$. For each policy, we convert the policy-specific candidate scores into a textual target output that lists candidates with their scores. We use sequence-generation supervision to align the training objective with the score-generation format used during inference. The policy-specific scorer is trained to align its generated output with this target:
\begin{equation}
\min_{\theta_{l,r}}
\sum_{(q,C(q),u)\in D}
\mathcal{L}_{\mathrm{gen}}
\left(
T_{l,r}(q),
\hat{T}_{l,r}(q)
\right),
\end{equation}
where $T_{l,r}(q)$ is the target scoring text for policy $\pi_{l,r}$, $\hat{T}_{l,r}(q)$ is the generated scoring text, and $\mathcal{L}_{\mathrm{gen}}$ is the sequence-generation loss.

Finally, we construct the support bank used by the fusion module. For each support instance of a preference scenario $\gamma$, we have the outputs of all trained policy-specific scorers and parse their outputs into score vectors. These vectors are normalized within the candidate set and stored together with the support ranking supervision:
\begin{equation}
B_{\gamma}=\{(q^{(k)},C^{(k)},z^{(k)},y^{(k)})\}_{k=1}^{M}.
\end{equation}
Here, $z^{(k)}$ contains the normalized score vectors produced by all view-policy scorers, and $y^{(k)}$ is used only for weight fitting on support instances.

\nop{\section{Evaluation}

We evaluate HARP on query-based CVE prioritization to assess its ranking effectiveness, robustness across prioritization contexts and backbone LLMs, and the contribution of its main components. We first describe the evaluation setting, then introduce baselines and metrics, and finally present the main results and ablation studies.}

\vspace{-5pt}
\section{Evaluation}
\label{sec:evaluation}
\vspace{-5pt}
We evaluate HARP across preference scenarios and backbone LLMs, then analyze the contribution of its components. We first describe the datasets and protocol, then baselines and results. Appendix~\ref{app:support_protocol} details the support-bank protocol and mixed-support control; Appendix~\ref{app:alpha_sensitivity} and Appendix~\ref{app:case_study} report parameter and case studies.

\vspace{-5pt}
\subsection{Datasets and Evaluation Setting}

We construct a query-based CVE prioritization benchmark from existing CVE records and exploit-related sources. Based on these sources, we build a vulnerability knowledge graph and generate query-candidate ranking instances for evaluation. Each instance contains a natural-language query, a candidate CVE set retrieved from the graph, and candidate-level vulnerability evidence.

To study preference-dependent ranking, cybersecurity practitioners from our organization’s IT/security team annotated ground-truth rankings under three evaluation scenarios: \emph{importance}, \emph{exploit\_first}, and \emph{patch\_order} (Appendix~\ref{app:preference_policy}). These scenarios define evaluation objectives and are distinct from the policy templates used for adapter training. Their guidelines were developed independently and did not ask annotators to reproduce any policy weighting scheme. For each query--candidate set, annotators ranked candidates using the scenario-specific guidelines and available vulnerability evidence, resolving disagreements through discussion to obtain a consensus ranking. Thus, the evaluation labels are independent of the policy-specific adapter supervision. Although some scenarios and policies share high-level objectives, rankings were based on the complete evidence and qualitative guidelines rather than predefined formulas or policy-specific rules. 

Each preference scenario has a separate support/train/dev/test split and a support bank $B_{\gamma}$ built only from its support split. At test time, HARP receives the query, graph-grounded candidate evidence, and historical supports from $B_{\gamma}$. The scoring prompt includes no scenario name, annotation guideline, or feature-weight vector, and the model does not predict the scenario identity. Fusion samples three supports to fit the view weights. A Mixed Support Bank control pools supports across scenarios (Appendix~\ref{app:support_protocol}). Splits are instance-based.

We evaluate GPT-OSS-20B, Llama3-8B, and Qwen2.5-7B-Instruct under each preference scenario using Precision@K, MAP@K, and MRR@K, where $K=\lceil 0.3\cdot|C(q)|\rceil$. Results are averaged over five independent runs, with HARP standard deviations reported in Appendix~\ref{app:run_variance}.

\nop{\subsection{Baselines}

We compare HARP with several representative baselines. \emph{LLM} directly prompts the base LLM to rank the candidate CVEs according to the query. \emph{LLM-SC} uses self-consistency by sampling multiple LLM-generated rankings and aggregating them into a final ranking. \emph{LLM-KG-RAG} retrieves and summarizes relevant evidence from the vulnerability knowledge graph and adds this information to the LLM prompt for ranking. \emph{Embedding Ranking} ranks candidates according to the embedding similarity between the query and each candidate's representation. We also include three single-view variants of our framework: \emph{Global}, \emph{Enterprise}, and \emph{User}, which use only the corresponding prioritization view for scoring without multi-view fusion. These baselines allow us to compare HARP against direct LLM reasoning, retrieval-augmented LLM reasoning, semantic similarity ranking, and non-fused single-view scoring.}

\vspace{-7pt}
\subsection{Baselines}
\vspace{-5pt}
We compare HARP with LLM-based and non-LLM baselines under their intended inputs. \emph{LLM} directly prompts the backbone to rank candidates; \emph{LLM-SC} aggregates multiple sampled rankings; \emph{LLM-KG-RAG} adds graph-retrieved evidence to the ranking prompt; and \emph{Embedding Ranking} uses query--candidate embedding similarity. We also include three single-view variants (\emph{Global}, \emph{Enterprise}, and \emph{User}) without multi-view fusion. As non-LLM comparisons, \emph{BM25} ranks candidates by lexical matching over textual query--candidate evidence, and \emph{Industrial Signals} builds a supervised ranker on structured vulnerability features that include CVSS severity, EPSS-style exploit likelihood, CISA KEV-style known-exploitation indicators, patch availability, attack vector, and related exposure cues. Thus, Industrial Signals uses the main industrial prioritization signals and learns how to combine them, which will serve as a traditional way for CVE ranking. 
%HARP differs from these baselines in two ways: it uses policy-conditioned multi-view scoring, and it adapts fusion with a historical support bank from the current preference scenario. 
%Prompting, BM25, and embedding baselines do not use such a bank; Industrial Signals uses structured feature supervision instead. Equal Weight and Mixed Support Bank (Section~\ref{sec:ablation}) then hold the HARP architecture fixed and isolate the contribution of preference-specific support supervision.

\begin{table*}[t]
\centering
\scriptsize
\setlength{\tabcolsep}{3pt}
\renewcommand{\arraystretch}{1.15}
\caption{Main ranking performance across non-LLM baselines, backbone LLMs, and preference scenarios. We report Precision@k, MAP@k, and MRR@k averaged over five runs. The best result for each metric within each backbone setting is highlighted in \textbf{bold}. HARP standard deviations are in Appendix~\ref{app:run_variance}.}
\label{tab:main_results}
\vspace{-5pt}
\resizebox{\textwidth}{!}{
\begin{tabular}{llccccccccc}
\toprule
\multirow{2}{*}{\textbf{Setting}} & \multirow{2}{*}{\textbf{Method}}
& \multicolumn{3}{c}{\textbf{Importance}}
& \multicolumn{3}{c}{\textbf{Exploit\_First}}
& \multicolumn{3}{c}{\textbf{Patch\_Order}} \\
\cmidrule(lr){3-5}\cmidrule(lr){6-8}\cmidrule(lr){9-11}
& & \textbf{P@k} & \textbf{MAP@k} & \textbf{MRR@k}
& \textbf{P@k} & \textbf{MAP@k} & \textbf{MRR@k}
& \textbf{P@k} & \textbf{MAP@k} & \textbf{MRR@k} \\
\midrule
\multirow{2}{*}{Non-LLM}
& Industrial Signals & 0.4945 & 0.3594 & 0.3304 & 0.5057 & 0.3672 & 0.3385 & 0.4966 & 0.3621 & 0.3327 \\
& BM25 & 0.4882 & 0.3783 & 0.3319 & 0.4566 & 0.3487 & 0.3066 & 0.4755 & 0.3703 & 0.3277 \\
\midrule
\multirow{8}{*}{GPT-OSS-20B}
& HARP & \textbf{0.5685} & \textbf{0.4598} & \textbf{0.4194} & \textbf{0.5758} & \textbf{0.4534} & \textbf{0.4123} & \textbf{0.5921} & \textbf{0.4833} & \textbf{0.4414} \\
& Global & 0.4700 & 0.4208 & 0.3541 & 0.4422 & 0.3946 & 0.3228 & 0.4634 & 0.4168 & 0.3495 \\
& Enterprise & 0.4852 & 0.4357 & 0.3709 & 0.4730 & 0.4167 & 0.3505 & 0.5117 & 0.4522 & 0.3876 \\
& User & 0.4659 & 0.4127 & 0.3490 & 0.4607 & 0.4125 & 0.3471 & 0.4749 & 0.4212 & 0.3607 \\
& LLM & 0.1943 & 0.1732 & 0.1371 & 0.2281 & 0.2036 & 0.1497 & 0.2261 & 0.2059 & 0.1651 \\
& LLM-SC & 0.4768 & 0.3693 & 0.3374 & 0.4829 & 0.3701 & 0.3389 & 0.4906 & 0.3839 & 0.3512 \\
& LLM-KG-RAG & 0.3032 & 0.2749 & 0.2246 & 0.3347 & 0.3050 & 0.2494 & 0.3014 & 0.2781 & 0.2250 \\
& Embedding & 0.4789 & 0.3618 & 0.3243 & 0.4883 & 0.3666 & 0.3284 & 0.4866 & 0.3704 & 0.3313 \\
\midrule
\multirow{8}{*}{Llama3-8B}
& HARP & \textbf{0.7466} & \textbf{0.6686} & \textbf{0.5871} & \textbf{0.7366} & \textbf{0.6597} & \textbf{0.5782} & \textbf{0.7449} & \textbf{0.6666} & \textbf{0.5859} \\
& Global & 0.6576 & 0.6302 & 0.5406 & 0.6720 & 0.6432 & 0.5443 & 0.6687 & 0.6380 & 0.5441 \\
& Enterprise & 0.6772 & 0.6409 & 0.5487 & 0.6768 & 0.6438 & 0.5471 & 0.6764 & 0.6471 & 0.5497 \\
& User & 0.6960 & 0.6634 & 0.5633 & 0.6739 & 0.6450 & 0.5434 & 0.6777 & 0.6432 & 0.5506 \\
& LLM & 0.4531 & 0.3800 & 0.3429 & 0.4283 & 0.3502 & 0.3162 & 0.4350 & 0.3602 & 0.3284 \\
& LLM-SC & 0.5331 & 0.4292 & 0.3920 & 0.5115 & 0.4002 & 0.3616 & 0.5160 & 0.4110 & 0.3734 \\
& LLM-KG-RAG & 0.6557 & 0.6198 & 0.5294 & 0.5888 & 0.5519 & 0.4732 & 0.6615 & 0.6275 & 0.5394 \\
& Embedding & 0.4793 & 0.3596 & 0.3198 & 0.4763 & 0.3553 & 0.3146 & 0.4785 & 0.3610 & 0.3197 \\
\midrule
\multirow{8}{*}{Qwen2.5-7B-Instruct}
& HARP & \textbf{0.8337} & \textbf{0.7376} & \textbf{0.6520} & \textbf{0.8336} & \textbf{0.7342} & \textbf{0.6488} & \textbf{0.8404} & \textbf{0.7417} & \textbf{0.6530} \\
& Global & 0.7282 & 0.6970 & 0.5866 & 0.7365 & 0.7052 & 0.5917 & 0.7246 & 0.6936 & 0.5842 \\
& Enterprise & 0.7447 & 0.7146 & 0.5981 & 0.7442 & 0.7137 & 0.5975 & 0.7508 & 0.7259 & 0.6109 \\
& User & 0.7565 & 0.7233 & 0.6106 & 0.7407 & 0.7067 & 0.5974 & 0.7686 & 0.7371 & 0.6229 \\
& LLM & 0.3587 & 0.2992 & 0.2551 & 0.3433 & 0.2828 & 0.2474 & 0.3425 & 0.2868 & 0.2486 \\
& LLM-SC & 0.5178 & 0.4073 & 0.3758 & 0.4944 & 0.3772 & 0.3491 & 0.5260 & 0.4155 & 0.3868 \\
& LLM-KG-RAG & 0.7192 & 0.6346 & 0.5709 & 0.6711 & 0.5784 & 0.5205 & 0.7252 & 0.6393 & 0.5723 \\
& Embedding & 0.4553 & 0.3404 & 0.3098 & 0.4438 & 0.3257 & 0.2953 & 0.4459 & 0.3306 & 0.2996 \\
\bottomrule
\vspace{-20pt}
\end{tabular}
}
\end{table*}

\nop{\subsection{Main Results}

Table~\ref{tab:main_results} reports the main ranking results across three backbone LLMs and three prioritization scenarios. HARP consistently outperforms direct LLM ranking, LLM-SC, LLM-KG-RAG, embedding-based ranking, and the single-view variants in almost all settings, showing that query-based CVE prioritization cannot be solved well by either a fixed scoring signal or a single prompting strategy. Direct LLM ranking performs poorly, and self-consistency improves it only partially, suggesting that repeated sampling does not address the missing implicit preference. LLM-KG-RAG benefits from graph-grounded evidence but still lags behind HARP, which indicates that access to vulnerability evidence alone is insufficient without preference-adaptive scoring. Embedding Ranking is competitive in some Precision@k results but is weaker on MAP@k and MRR@k, showing that semantic similarity does not reliably capture remediation priority. The single-view variants also perform worse than the full model, confirming that global, enterprise, and user views provide complementary signals and need to be combined adaptively for different prioritization scenarios.}
\vspace{-5pt}
\subsection{Main Results}

Table~\ref{tab:main_results} reports five-run averages across non-LLM baselines, three backbone LLMs, and three preference scenarios. 
HARP uses multi-view scoring and Preference-adaption that makes our method provide better performance, while other method like prompting, BM25, and embedding baselines do not. 
HARP consistently outperforms LLM prompting, LLM-SC, LLM-KG-RAG, embedding ranking, single-view variants, and the non-LLM baselines. Direct LLM ranking is weak and self-consistency helps only partially. LLM-KG-RAG benefits from graph evidence but still lags behind HARP. Embedding Ranking is competitive on some Precision@$K$ scores but weaker on MAP@$K$ and MRR@$K$. BM25 remains below HARP, showing that lexical matching alone is insufficient. Industrial Signals is stronger than several prompting baselines but still underperforms HARP, even though it directly consumes industrial prioritization signals as explicit feature vectors. The single-view variants also underperform the full model. While a few MAP@$K$ margins over the strongest single view are smaller than the gaps to prompting baselines, HARP remains best across P@$K$/MAP@$K$/MRR@$K$ jointly, and Appendix~\ref{app:run_variance} shows that these improvements are stable relative to run-to-run variation. Within-system evidence that scenario-matched supports matter is reported in Section~\ref{sec:ablation}.

We provide further analysis in the appendix. Appendix~\ref{app:run_variance} reports run-to-run standard deviations for HARP, Appendix~\ref{app:alpha_sensitivity} reports a sensitivity study on the interpolation coefficient $\alpha$, Appendix~\ref{app:support_protocol} details the support-bank protocol and mixed-support control, and Appendix~\ref{app:case_study} presents cases analyzing how HARP combines view-specific rankings.

\nop{\subsection{Ablation Study}

Table~\ref{tab:ablation_results} reports the ablation results. We compare HARP with five variants: \emph{Fusion}, which removes the uniform multi-view regularization term and uses only the adaptive fused score; \emph{Equal Weight}, which assigns the same fusion weight to all views; \emph{Random Weight}, which replaces the fitted fusion weights with random weights; \emph{No Top-K Router}, which disables the consensus shortcut that directly preserves the shared top-k ordering when all views agree; and \emph{Random Policy}, which randomly selects policies instead of using query-conditioned policy selection. The results show that each component contributes to the final performance. Fusion generally performs worse than HARP, especially in the exploit-first and patch-order contexts, indicating that the uniform branch helps preserve evidence from all views. Equal Weight is consistently weaker than HARP, while Random Weight causes a much larger drop, confirming that view contributions should be estimated rather than fixed or random. No Top-K Consensus also underperforms the full model, suggesting that preserving cross-view agreement among top candidates improves ranking stability. Random Policy is generally weaker than HARP, demonstrating that query-conditioned policy selection is important for matching the scoring behavior to the implicit prioritization preference.}

\begin{table*}[t]
\centering
\scriptsize
\setlength{\tabcolsep}{3pt}
\renewcommand{\arraystretch}{1.15}
\caption{Ablation study across backbone LLMs and preference scenarios. Mixed Support Bank pools supports from all preference scenarios without using the evaluated scenario identity. We report Precision@k, MAP@k, and MRR@k. The best result for each metric within each setting is highlighted in \textbf{bold}.}
\vspace{-5pt}
\label{tab:ablation_results}
\resizebox{\textwidth}{!}{
\begin{tabular}{llccccccccc}
\toprule
\multirow{2}{*}{\textbf{Backbone LLM}} & \multirow{2}{*}{\textbf{Method}}
& \multicolumn{3}{c}{\textbf{Importance}}
& \multicolumn{3}{c}{\textbf{Exploit\_First}}
& \multicolumn{3}{c}{\textbf{Patch\_Order}} \\
\cmidrule(lr){3-5}\cmidrule(lr){6-8}\cmidrule(lr){9-11}
& & \textbf{P@k} & \textbf{MAP@k} & \textbf{MRR@k}
& \textbf{P@k} & \textbf{MAP@k} & \textbf{MRR@k}
& \textbf{P@k} & \textbf{MAP@k} & \textbf{MRR@k} \\
\midrule
\multirow{7}{*}{GPT-OSS-20B}
& HARP & \textbf{0.5685} & \textbf{0.4598} & \textbf{0.4194} & \textbf{0.5758} & \textbf{0.4534} & \textbf{0.4123} & \textbf{0.5921} & \textbf{0.4833} & \textbf{0.4414} \\
& Mixed Support Bank & 0.5340 & 0.4395 & 0.3985 & 0.5175 & 0.4180 & 0.3765 & 0.5585 & 0.4580 & 0.4175 \\
& No Regularization & 0.5601 & 0.4526 & 0.4128 & 0.5204 & 0.4051 & 0.3706 & 0.5605 & 0.4431 & 0.4076 \\
& Equal Weight & 0.5317 & 0.4380 & 0.3974 & 0.5192 & 0.4197 & 0.3776 & 0.5608 & 0.4604 & 0.4190 \\
& Random Weight & 0.4056 & 0.3109 & 0.2926 & 0.4115 & 0.3204 & 0.2963 & 0.4106 & 0.3071 & 0.2881 \\
& No Top-K Consensus & 0.5344 & 0.4407 & 0.3996 & 0.4787 & 0.3812 & 0.3458 & 0.5281 & 0.4223 & 0.3864 \\
& Random Policy & 0.4920 & 0.3985 & 0.3652 & 0.4728 & 0.3761 & 0.3410 & 0.5015 & 0.4082 & 0.3755 \\
\midrule
\multirow{7}{*}{Llama3-8B}
& HARP & \textbf{0.7466} & \textbf{0.6686} & \textbf{0.5871} & \textbf{0.7366} & \textbf{0.6597} & \textbf{0.5782} & \textbf{0.7449} & \textbf{0.6666} & \textbf{0.5859} \\
& Mixed Support Bank & 0.7112 & 0.6465 & 0.5640 & 0.7088 & 0.6495 & 0.5632 & 0.7060 & 0.6505 & 0.5710 \\
& No Regularization & 0.7186 & 0.6404 & 0.5675 & 0.7160 & 0.6380 & 0.5681 & 0.7093 & 0.6295 & 0.5557 \\
& Equal Weight & 0.7095 & 0.6489 & 0.5657 & 0.7104 & 0.6528 & 0.5649 & 0.7045 & 0.6528 & 0.5736 \\
& Random Weight & 0.1858 & 0.1367 & 0.1126 & 0.1899 & 0.1378 & 0.1122 & 0.1718 & 0.1287 & 0.1061 \\
& No Top-K Consensus & 0.6795 & 0.6206 & 0.5490 & 0.6846 & 0.6267 & 0.5530 & 0.6668 & 0.6115 & 0.5362 \\
& Random Policy & 0.7052 & 0.6285 & 0.5510 & 0.6684 & 0.5912 & 0.5208 & 0.6987 & 0.6320 & 0.5524 \\
\midrule
\multirow{7}{*}{Qwen2.5-7B-Instruct}
& HARP & \textbf{0.8337} & \textbf{0.7376} & \textbf{0.6520} & \textbf{0.8336} & \textbf{0.7342} & \textbf{0.6488} & \textbf{0.8404} & \textbf{0.7417} & \textbf{0.6530} \\
& Mixed Support Bank & 0.7725 & 0.7148 & 0.6195 & 0.7680 & 0.7125 & 0.6160 & 0.7835 & 0.7265 & 0.6275 \\
& No Regularization & 0.8028 & 0.7018 & 0.6255 & 0.8165 & 0.7139 & 0.6347 & 0.8290 & 0.7299 & 0.6442 \\
& Equal Weight & 0.7697 & 0.7163 & 0.6179 & 0.7651 & 0.7147 & 0.6143 & 0.7809 & 0.7282 & 0.6261 \\
& Random Weight & 0.1584 & 0.1160 & 0.0850 & 0.1568 & 0.1186 & 0.0876 & 0.1500 & 0.1115 & 0.0838 \\
& No Top-K Consensus & 0.7383 & 0.6813 & 0.5918 & 0.7461 & 0.6919 & 0.5985 & 0.7565 & 0.7023 & 0.6051 \\
& Random Policy & 0.7864 & 0.6885 & 0.6120 & 0.7620 & 0.6588 & 0.5885 & 0.7912 & 0.6985 & 0.6108 \\
\bottomrule
\vspace{-20pt}
\end{tabular}
}
\end{table*}

\vspace{-10pt}
\subsection{Ablation Study}
\label{sec:ablation}
\vspace{-5pt}
Table~\ref{tab:ablation_results} reports the ablation results. We compare HARP with Mixed Support Bank, No Regularization, Equal Weight, Random Weight, No Top-K Consensus, and Random Policy. Mixed Support Bank keeps the same adapters and fusion procedure but pools supports across preference scenarios and samples without using the evaluated scenario identity; scores drop relative to HARP to a level close to Equal Weight and generally below No Regularization, showing that scenario-matched support supervision is an important adaptation signal. No Regularization generally performs worse than HARP, especially in the exploit-first and patch-order preference scenarios, indicating that the uniform branch helps preserve complementary evidence across views. Equal Weight consistently underperforms HARP, while Random Weight causes a much larger drop, confirming that view contributions should be estimated rather than fixed or random. No Top-K Consensus also underperforms the full model, suggesting that preserving cross-view agreement among top candidates improves ranking stability. Random Policy is generally weaker than HARP, demonstrating the importance of query-conditioned policy selection.

\vspace{-6pt}
\section{Related Work}
\vspace{-3pt}

\subsection{Vulnerability Prioritization and Ranking}

Vulnerability prioritization is closely related to exploit prediction and remediation planning. Prior work shows that exploit-based signals, social evidence, patch availability, and operational constraints are important for improving remediation decisions and risk assessment~\cite{jacobs2020remediation,chen2019twitter,olswang2022patching,lacombe2025attacker}. These studies motivate evidence-aware vulnerability ranking, but they mainly focus on exploit prediction, patch ordering, or fixed risk criteria. General ranking and neural embedding retrieval models further provide generic ranking tools~\cite{cao2007listnet,burges2006lambdarank,nogueira2020documentranking}. However, they typically assume that the target ranking criterion is fixed or explicitly supervised. In contrast, HARP studies CVE prioritization under an existing preference scenario, where preference is carried by a scenario-specific bank of past labeled supports rather than by the query or a fixed scoring rule.
\vspace{-8pt}
\subsection{Modular Adaptation}
\vspace{-2pt}

Retrieval-augmented and evidence-grounded language models improve generation by incorporating external information~\cite{karpukhin2020dpr,khattab2020colbert,lewis2020rag,guu2020realm}. However, grounding alone does not specify which preference scenario should guide ranking. Parameter-efficient adaptation methods such as adapters and LoRA provide lightweight modules for specializing model behavior without full retraining~\cite{houlsby2019adapter,hu2022lora}. Knowledge editing and related modular-update methods further study localized parameter changes for factual or behavioral control~\cite{mitchell2022mend,mitchell2022serac,dai2022knowledge,zhong2023mquake,hartvigsen2023grace,yu2024melo,wang2024wise,li2025elder,zhang2026loka,liu2026representation,zhang2026mind,zou2023repe,deng2025unke,jung2025come}, but they mainly target factual updating or task adaptation. HARP instead uses policy-specific lightweight adapters for multi-view CVE scoring and combines them through adaptive score fusion; we treat knowledge editing only as related motivation for modular control, not as the technical basis of our method.

\vspace{-5pt}
\section{Conclusion}
\vspace{-5pt}
In this paper, we studied query-based CVE prioritization under preference scenarios that are often implicit and difficult to specify in prompts. We proposed HARP, a graph-grounded multi-view framework that retrieves vulnerability evidence, performs policy-conditioned scoring, and adapts view fusion using scenario-matched historical supports without requiring an explicit textual preference description. We also constructed a benchmark with policy-specific training supervision and practitioner-annotated rankings under three preference scenarios. Experiments show that HARP outperforms multiple standard ranking baselines as a full system.

\section{Limitations}

This work has several limitations. Our benchmark focuses on three preference scenarios and English triage queries built from public CVE evidence; extending to additional languages, proprietary asset inventories, or organization-specific remediation workflows is left for future work. The current implementation freezes the backbone LLM and trains only lightweight policy adapters, so the effect of jointly updating more backbone parameters is not explored. Inference uses multiple view-wise scoring calls, and wall-clock cost may vary with candidate-set size and hardware. 

\section{Ethical Considerations}

This work is intended to support defensive vulnerability management by helping analysts prioritize CVEs for remediation. The framework does not generate exploits or provide instructions for exploiting vulnerabilities; instead, it uses vulnerability evidence to improve ranking decisions. However, any system that prioritizes high-risk or actively exploitable vulnerabilities could potentially be misused if deployed without appropriate access restrictions. Therefore, practical use should be limited to authorized security workflows and should follow responsible disclosure and organizational security policies. The model may also inherit bias or incompleteness from public vulnerability records, exploit databases, or patch information, which could lead to incorrect prioritization. Human oversight remains necessary, especially for high-impact remediation decisions, and the system should be used as a decision-support tool rather than an autonomous replacement for security analysts.

\bibliography{custom}

@misc{cvss31,
  title  = {Common Vulnerability Scoring System Version 3.1: Specification Document},
  author = {{FIRST.Org, Inc.}},
  year   = {2019},
  url       = {https://www.first.org/cvss/v3-1/cvss-v31-specification_r1.pdf}

}

@inproceedings{epss2023,
  title     = {Enhancing Vulnerability Prioritization: Data-Driven Exploit Predictions with Community-Driven Insights},
  author    = {Jacobs, Jay and Romanosky, Sasha and Suciu, Octavian and Edwards, Benjamin and Sarabi, Armin},
  booktitle = {IEEE European Symposium on Security and Privacy Workshops (EuroS\&PW)},
  year      = {2023}
}

@techreport{ssvc,
  title       = {Prioritizing Vulnerability Response: A Stakeholder-Specific Vulnerability Categorization (Version 2.0)},
  author      = {Spring, Jonathan M. and Householder, Allen D. and Hatleback, Eric and Manion, Art and Oliver, Madison and Sarvepalli, Vijay S. and Tyzenhaus, Laurie and Yarbrough, Charles G.},
  institution = {Carnegie Mellon University Software Engineering Institute},
  year        = {2021}
}

@inproceedings{bozorgi2010beyond,
  title     = {Beyond Heuristics: Learning to Classify Vulnerabilities and Predict Exploits},
  author    = {Bozorgi, Mehran and Saul, Lawrence K. and Savage, Stefan and Voelker, Geoffrey M.},
  booktitle = {ACM SIGKDD Conference on Knowledge Discovery and Data Mining (KDD)},
  year      = {2010}
}

@inproceedings{suciu2022expected,
  title     = {Expected Exploitability: Predicting the Development of Functional Vulnerability Exploits},
  author    = {Suciu, Octavian and Nelson, Connor and Lyu, Zhuoer and Bao, Tiffany and Dumitra{\c{s}}, Tudor},
  booktitle = {USENIX Security Symposium (USENIX Security)},
  year      = {2022}
}

@inproceedings{wudi2022diffcvss,
  title     = {{OS}-Aware Vulnerability Prioritization via Differential Severity Analysis},
  author    = {Wu, Qiushi and Xiao, Yue and Liao, Xiaojing and Lu, Kangjie},
  booktitle = {USENIX Security Symposium (USENIX Security)},
  year      = {2022}
}

@inproceedings{burges2005ranknet,
  title     = {Learning to Rank Using Gradient Descent},
  author    = {Burges, Christopher J. C. and Shaked, Tal and Renshaw, Erin and Lazier, Ari and Deeds, Matt and Hamilton, Nicole and Hullender, Gregory N.},
  booktitle = {International Conference on Machine Learning (ICML)},
  year      = {2005}
}

@techreport{burges2010lambdamart,
  title       = {From RankNet to LambdaRank to LambdaMART: An Overview},
  author      = {Burges, Christopher J. C.},
  institution = {Microsoft Research},
  year        = {2010}
}

@inproceedings{karpukhin2020dpr,
  title     = {Dense Passage Retrieval for Open-Domain Question Answering},
  author    = {Karpukhin, Vladimir and Oguz, Barlas and Min, Sewon and Lewis, Patrick and Wu, Ledell and Edunov, Sergey and Chen, Danqi and Yih, Wen-tau},
  booktitle = {Conference on Empirical Methods in Natural Language Processing (EMNLP)},
  year      = {2020}
}

@inproceedings{khattab2020colbert,
  title     = {{ColBERT}: Efficient and Effective Passage Search via Contextualized Late Interaction over {BERT}},
  author    = {Khattab, Omar and Zaharia, Matei},
  booktitle = {International ACM SIGIR Conference on Research and Development in Information Retrieval (SIGIR)},
  year      = {2020}
}

@inproceedings{lewis2020rag,
  title     = {Retrieval-Augmented Generation for Knowledge-Intensive {NLP} Tasks},
  author    = {Lewis, Patrick and Perez, Ethan and Piktus, Aleksandra and Petroni, Fabio and Karpukhin, Vladimir and Goyal, Naman and K{\"u}ttler, Heinrich and Lewis, Mike and Yih, Wen-tau and Rockt{\"a}schel, Tim and Riedel, Sebastian and Kiela, Douwe},
  booktitle = {Advances in Neural Information Processing Systems (NeurIPS)},
  year      = {2020}
}

@inproceedings{guu2020realm,
  title     = {{REALM}: Retrieval-Augmented Language Model Pre-Training},
  author    = {Guu, Kelvin and Lee, Kenton and Tung, Zora and Pasupat, Panupong and Chang, Ming-Wei},
  booktitle = {International Conference on Machine Learning (ICML)},
  year      = {2020}
}

@inproceedings{houlsby2019adapter,
  title     = {Parameter-Efficient Transfer Learning for {NLP}},
  author    = {Houlsby, Neil and Giurgiu, Andrei and Jastrzebski, Stanislaw and Morrone, Bruna and De Laroussilhe, Quentin and Gesmundo, Andrea and Attariyan, Mona and Gelly, Sylvain},
  booktitle = {International Conference on Machine Learning (ICML)},
  year      = {2019}
}

@inproceedings{hu2022lora,
  title     = {{LoRA}: Low-Rank Adaptation of Large Language Models},
  author    = {Hu, Edward J. and Shen, Yelong and Wallis, Phillip and Allen-Zhu, Zeyuan and Li, Yuanzhi and Wang, Shean and Wang, Lu and Chen, Weizhu},
  booktitle = {International Conference on Learning Representations (ICLR)},
  year      = {2022}
}

@inproceedings{meng2022rome,
  title     = {Locating and Editing Factual Associations in {GPT}},
  author    = {Meng, Kevin and Bau, David and Andonian, Alex and Belinkov, Yonatan},
  booktitle = {Advances in Neural Information Processing Systems (NeurIPS)},
  year      = {2022}
}

@inproceedings{meng2023memit,
  title     = {Mass-Editing Memory in a Transformer},
  author    = {Meng, Kevin and Sen Sharma, Arnab and Andonian, Alex and Belinkov, Yonatan and Bau, David},
  booktitle = {International Conference on Learning Representations (ICLR)},
  year      = {2023}
}

@article{wang2024knowledgeediting,
  title   = {Knowledge Editing for Large Language Models: A Survey},
  author  = {Wang, Song and Zhu, Yaochen and Liu, Haochen and Zheng, Zaiyi and Chen, Chen and Li, Jundong},
  journal = {ACM Computing Surveys},
  year    = {2024}
}

@article{zhang2026loka,
  title     = {Resolving Editing-Unlearning Conflicts: A Knowledge Codebook Framework for Large Language Model Updating},
  author    = {Zhang, Binchi and Chen, Zhengzhang and Zheng, Zaiyi and Li, Jundong and Chen, Haifeng},
  journal = {arXiv:2502.00158},
  year      = {2025}
}

@article{liu2026representation,
  title     = {Representation Interventions Enable Lifelong Unstructured Knowledge Control},
  author    = {Liu, Xuyuan and Chen, Shengyu and Dong, Xinshuai and Liu, Yanchi and Zhao, Xujiang and Wang, Haoyu and Yan, Yujun and Chen, Haifeng and Chen, Zhengzhang},
  journal = {arXiv:2511.20892},
  year      = {2025}
}

@article{zhang2026mind,
  title     = {Mind the Gap in Cultural Alignment: Task-Aware Culture Management for Large Language Models},
  author    = {Zhang, Binchi and Zhao, Xujiang and Li, Jundong and Chen, Haifeng and Chen, Zhengzhang},
  journal = {arXiv:2602.22475},
  year      = {2026}
}

@inproceedings{li2017patches,
author = {Li, Frank and Paxson, Vern},
title = {A Large-Scale Empirical Study of Security Patches},
year = {2017},
booktitle = {ACM Conference on Computer and Communications Security (CCS)},
}

@inproceedings{desmale2023firehose,
  title     = {No One Drinks From the Firehose: How Organizations Filter and Prioritize Vulnerability Information},
  author    = {de Smale, Stephanie and van Dijk, Rik and Bouwman, Xander B. and van der Ham, Jeroen and van Eeten, Michel},
  booktitle = {IEEE Symposium on Security and Privacy (IEEE S\&P)},
  year      = {2023}
}

@article{allodi2014risk,
  title   = {Comparing Vulnerability Severity and Exploits Using Case-Control Studies},
  author  = {Allodi, Luca and Massacci, Fabio},
  journal = {ACM TISSEC},
  year    = {2014}
}

@inproceedings{sabottke2015vulnerability,
  title     = {Vulnerability Disclosure in the Age of Social Media: Exploiting Twitter for Predicting Real-World Exploits},
  author    = {Sabottke, Carl and Suciu, Octavian and Dumitra{\c{s}}, Tudor},
  booktitle = {USENIX Security Symposium (USENIX Security)},
  year      = {2015}
}

@article{jacobs2020remediation,
  title   = {Improving Vulnerability Remediation Through Better Exploit Prediction},
  author  = {Jacobs, Jay and Romanosky, Sasha and Adjerid, Idris and Baker, Wade},
  journal = {Journal of Cybersecurity},
  year    = {2020}
}

@inproceedings{chen2019twitter,
  title     = {Using Twitter to Predict When Vulnerabilities will be Exploited},
  author    = {Chen, Haipeng and Liu, Rui and Park, Noseong and Subrahmanian, V. S.},
  booktitle = {ACM SIGKDD Conference on Knowledge Discovery and Data Mining (KDD)},
  year      = {2019}
}

@article{olswang2022patching,
  title   = {Prioritizing Vulnerability Patches in Large Networks},
  author  = {Amir Olswang and Tom Gonda and Rami Puzis and Guy Shani and Bracha Shapira and Noam Tractinsky},
  journal = {Expert Systems with Applications},
  year    = {2022}
}

@inproceedings{lacombe2025attacker,
  title     = {Attacker Control and Bug Prioritization},
  author    = {Lacombe, Guillaume and Bardin, S{\'e}bastien},
  booktitle = {USENIX Security Symposium (USENIX Security)},
  year      = {2025}
}

@inproceedings{cao2007listnet,
  title     = {Learning to Rank: From Pairwise Approach to Listwise Approach},
  author    = {Cao, Zhe and Qin, Tao and Liu, Tie-Yan and Tsai, Ming-Feng and Li, Hang},
  booktitle = {International Conference on Machine Learning (ICML)},
  year      = {2007}
}

@inproceedings{burges2006lambdarank,
  title     = {Learning to Rank with Nonsmooth Cost Functions},
  author    = {Burges, Christopher J. C. and Ragno, Robert and Le, Quoc V.},
  booktitle = {Advances in Neural Information Processing Systems (NeurIPS)},
  year      = {2006}
}

@inproceedings{nogueira2020documentranking,
  title     = {Document Ranking with a Pretrained Sequence-to-Sequence Model},
  author    = {Nogueira, Rodrigo and Jiang, Zhiying and Lin, Jimmy},
  booktitle = {Conference on Empirical Methods in Natural Language Processing (EMNLP)},
  year      = {2020}
}

@inproceedings{mitchell2022mend,
  title     = {Fast Model Editing at Scale},
  author    = {Mitchell, Eric and Lin, Charles and Bosselut, Antoine and Finn, Chelsea and Manning, Christopher D.},
  booktitle = {International Conference on Learning Representations (ICLR)},
  year      = {2022}
}

@inproceedings{mitchell2022serac,
  title     = {Memory-Based Model Editing at Scale},
  author    = {Mitchell, Eric and Lin, Charles and Bosselut, Antoine and Manning, Christopher D. and Finn, Chelsea},
  booktitle = {International Conference on Machine Learning (ICML)},
  year      = {2022}
}

@inproceedings{dai2022knowledge,
  title     = {Knowledge Neurons in Pretrained Transformers},
  author    = {Dai, Damai and Dong, Li and Hao, Yaru and Sui, Zhifang and Chang, Baobao and Wei, Furu},
  booktitle = {Annual Meeting of the Association for Computational Linguistics (ACL)},
  year      = {2022}
}

@inproceedings{zhong2023mquake,
  title     = {{MQuAKE}: Assessing Knowledge Editing in Language Models via Multi-Hop Questions},
  author    = {Zhong, Zexuan and Wu, Zhengxuan and Manning, Christopher D. and Potts, Christopher and Chen, Danqi},
  booktitle = {Conference on Empirical Methods in Natural Language Processing (EMNLP)},
  year      = {2023}
}

@inproceedings{hartvigsen2023grace,
  title     = {Aging with {GRACE}: Lifelong Model Editing with Discrete Key-Value Adaptors},
  author    = {Hartvigsen, Thomas and Sankaranarayanan, Swami and Palangi, Hamid and Kim, Yoon and Ghassemi, Marzyeh},
  booktitle = {Advances in Neural Information Processing Systems (NeurIPS)},
  year      = {2023}
}

@inproceedings{wang2024wise,
  title     = {{WISE}: Rethinking the Knowledge Memory for Lifelong Model Editing of Large Language Models},
  author    = {Wang, Peng and Li, Zexi and Zhang, Ningyu and Xu, Ziwen and Yao, Yunzhi and Jiang, Yong and Xie, Pengjun and Huang, Fei and Chen, Huajun},
  booktitle = {Advances in Neural Information Processing Systems (NeurIPS)},
  year      = {2024}
}

@article{qwen25,
  title     = {Qwen2.5 Technical Report},
  author    = {Yang, An and Yang, Baosong and Zhang, Beichen and Hui, Binyuan and Zheng, Bo and Yu, Bowen and Li, Chengyuan and Liu, Dayiheng and Huang, Fei and Wei, Haoran and Lin, Huan and Yang, Jian and Tu, Jianhong and Zhang, Jianwei and Yang, Jianxin and Yang, Jiaxi and Zhou, Jingren and Lin, Junyang and Dang, Kai and Lu, Keming and Bao, Keqin and Yang, Kexin and Yu, Le and Li, Mei and Xue, Mingfeng and Zhang, Pei and Zhu, Qin and Men, Rui and Lin, Runji and Li, Tianhao and Tang, Tianyi and Xia, Tingyu and Ren, Xingzhang and Ren, Xuancheng and Fan, Yang and Su, Yang and Zhang, Yichang and Wan, Yu and Liu, Yuqiong and Cui, Zeyu and Zhang, Zhenru and Qiu, Zihan},
  journal = {arXiv:2412.15115},
  year      = {2024}
}

@inproceedings{adamw,
  title     = {Decoupled Weight Decay Regularization},
  author    = {Loshchilov, Ilya and Hutter, Frank},
  booktitle = {International Conference on Learning Representations (ICLR)},
  year      = {2019}
}

@article{llama3herd,
  title     = {The Llama 3 Herd of Models},
  author    = {Grattafiori, Aaron and Dubey, Abhimanyu and Jauhri, Abhinav and Pandey, Abhinav and Kadian, Abhishek and Al-Dahle, Ahmad and Letman, Aiesha and Mathur, Akhil and Schelten, Alan and Vaughan, Alex and Yang, Amy and Fan, Angela and Goyal, Anirudh and Hartshorn, Anthony and Yang, Aobo and Mitra, Archi and Sravankumar, Archie and Korenev, Artem and Hinsvark, Arthur and Rao, Arun and Zhang, Aston and Rodriguez, Aurelien and Gregerson, Austen and Spataru, Ava and Roziere, Baptiste and Biron, Bethany and Tang, Binh and Chern, Bobbie and Caucheteux, Charlotte and Nayak, Chaya and Bi, Chloe and Marra, Chris and McConnell, Chris and Keller, Christian and Touret, Christophe and Wu, Chunyang and Wong, Corinne and Ferrer, Cristian Canton and Nikolaidis, Cyrus and Allonsius, Damien and Song, Daniel and Pintz, Danielle and Livshits, David and Wyatt, Danny and Esiobu, David and Choudhary, Dhruv and Mahajan, Dhruv and Garcia-Olano, Diego and Perino, Diego and Hupkes, Dieuwke and Lakomkin, Egor and AlBadawy, Ehab and Lobanova, Elina and Dinan, Emily and Smith, Eric Michael and Radenovic, Filip and Zhang, Frank and Synnaeve, Gabriel and Lee, Gabrielle and Anderson, Georgia Lewis and Thattai, Govind and Nail, Graeme and Mialon, Gr{\'e}goire and Pang, Guan and Cucurell, Guillem and Nguyen, Hailey and Korevaar, Hannah and Xu, Hu and Touvron, Hugo and Zarov, Iliyan and Ibarra, Imanol Arrieta and Kloumann, Isabel and Misra, Ishan and Evtimov, Ivan and Copet, Jade and Lee, Jaewon and Geffert, Jan and Vranes, Jana and Park, Jason and Mahadeokar, Jay and Shah, Jeet and van der Linde, Jelmer and Billock, Jennifer and Hong, Jenny and Lee, Jenya and Fu, Jeremy and Chi, Jianfeng and Huang, Jianyu and Liu, Jiawen and Wang, Jie and Yu, Jiecao and Bitton, Joanna and Spisak, Joe and Park, Jongsoo and Rocca, Joseph and Johnstun, Joshua and Saxe, Joshua and Jia, Junteng and Alwala, Kalyan Vasuden and Prasad, Karthik and Upasani, Kartikeya and Plawiak, Kate and Li, Ke and Heafield, Kenneth and Stone, Kevin and El-Arini, Khalid and Iyer, Krithika and Malik, Kshitiz and Chiu, Kuenley and Bhalla, Kunal and Rantala-Yeary, Lauren and van der Maaten, Laurens and Chen, Lawrence and Tan, Liang and Jenkins, Liz and Martin, Louis and Madaan, Lovish and Malo, Lubo and Blecher, Lukas and Landzaat, Lukas and de Oliveira, Madeline Muzzi and Muzzi, Madian Khabsa and Xia, Manling and Mavrinac, Manohar and Aldeen, Mansheej Paul and Zvyagina, Mara and Osama, Mark and Kambadur, Melanie and Lewis, Mike and Si, Min and Singh, Mitesh Kumar and Hassan, Mona and Goyal, Naman and Torabi, Narjes and Bashlykov, Nikolay and Bogoychev, Nikolay and Chatterji, Niladri and Duchenne, Olivier and Çelebi, Onur and Alrassy, Ousman and Zhang, Pengchuan and Li, Pengwei and Vasic, Petar and Weng, Peter and Bhargava, Prajjwal and Dubal, Pratik and Krishnan, Praveen and Koura, Punit Singh and Xu, Puxin and He, Qing and Dong, Qingxiao and Srinivasan, Ragavan and Ganapathy, Raj and Calderer, Ramon and Cabral, Ricardo Silveira and Stojnic, Robert and Raileanu, Roberta and Girdhar, Rohit and Patel, Rohit and Sauvestre, Romain and Polidoro, Rosario and Sumbaly, Roshan and Taylor, Ross and Silva, Ruan and Hou, Rui and Wang, Rui and Hosseini, Saghar and Chennabasappa, Sahana and Singh, Sanjay and Bell, Sean and Kim, Seohyun Sonia and Edunov, Sergey and Nie, Shaoliang and Narang, Sharan and Raparthy, Sharath and Shen, Sheng and Wan, Shengye and Bhosale, Shruti and Zhang, Shun and Vandenhende, Simon and Batra, Soumya and Whitman, Spencer and Sootla, Sten and Collot, Stephane and Gururangan, Suchin and Borodinsky, Sydney and Herman, Tamar and Fowler, Tara and Sheasha, Tarek and Georgiou, Thomas and Scialom, Thomas and Speckbacher, Tobias and Mihaylov, Todor and Xiao, Tong and Karn, Ujjwal and Goswami, Vedanuj and Gupta, Vibhor and Ramanathan, Vignesh and Kerkez, Viktor and Gonguet, Vincent and Do, Vinh Quang and Vogeti, Vish and Albiero, Vitor and Petrovic, Vladan and Chu, Weiwei and Xiong, Wenhan and Fu, Wenyin and Meers, Whitney and Martinet, Xavier and Wang, Xiaodong and Tan, Xiaoqing Ellen and Xia, Xinfeng and Xie, Xuchao and Jia, Xuewei and Wang, Xueyang and Goldschlag, Yaelle and Gaur, Yashesh and Babaei, Yasmine and Wen, Yi and Song, Yiwen and Zhang, Yuchen and Li, Yue and Mao, Yuning and Coudert, Zacharie Delpierre and Yan, Zheng and Chen, Zhengxing and Papakipos, Zoe and Singh, Aaditya and Grattafiori, Aaron and Dubey, Abhimanyu and Jauhri, Abhinav},
  journal = {arXiv:2407.21783},
  year      = {2024}
}

@inproceedings{host-etal-2023-constructing,
    title = "Constructing a Knowledge Graph from Textual Descriptions of Software Vulnerabilities in the National Vulnerability Database",
    author = "H{\o}st, Anders  and
      Lison, Pierre  and
      Moonen, Leon",
    booktitle = "Nordic Conference on Computational Linguistics (NoDaLiDa)",
    month = may,
    year = "2023",
}

@inproceedings{wu2024reft,
  title     = {{ReFT}: Representation Finetuning for Language Models},
  author    = {Wu, Zhengxuan and Arora, Aryaman and Wang, Zheng and Geiger, Atticus and Jurafsky, Dan and Manning, Christopher D. and Potts, Christopher},
  booktitle = {Advances in Neural Information Processing Systems (NeurIPS)},
  year      = {2024}
}

@inproceedings{yu2024melo,
  title     = {{MELO}: Enhancing Model Editing with Neuron-Indexed Dynamic {LoRA}},
  author    = {Yu, Lang and Chen, Qin and Zhou, Jie and He, Liang},
  booktitle = {AAAI Conference on Artificial Intelligence (AAAI)},
  year      = {2024}
}

@inproceedings{li2025elder,
  title     = {{ELDER}: Enhancing Lifelong Model Editing with Mixture-of-{LoRA}},
  author    = {Li, Jiaang and Wang, Quan and Wang, Zhongnan and Zhang, Yongdong and Mao, Zhendong},
  booktitle = {AAAI Conference on Artificial Intelligence (AAAI)},
  year      = {2025}
}

@inproceedings{deng2025unke,
  title     = {Everything is Editable: Extend Knowledge Editing to Unstructured Data in Large Language Models},
  author    = {Deng, Jingcheng and Wei, Zihao and Pang, Liang and Ding, Hanxing and Shen, Huawei and Cheng, Xueqi},
  booktitle = {International Conference on Learning Representations (ICLR)},
  year      = {2025}
}

@inproceedings{jung2025come,
  title     = {{CoME}: An Unlearning-Based Approach to Conflict-Free Model Editing},
  author    = {Jung, Dahyun and Seo, Jaehyung and Lee, Jaewook and Park, Chanjun and Lim, Heuiseok},
  booktitle = {Annual Conference of the North American Chapter of the Association for Computational Linguistics (NAACL)},
  year      = {2025}
}

@article{zou2023repe,
  title   = {Representation Engineering: A Top-Down Approach to {AI} Transparency},
  author  = {Zou, Andy and Phan, Long and Chen, Sarah and Campbell, James and Guo, Phillip and Ren, Richard and Pan, Alexander and Yin, Xuwang and Mazeika, Mantas and Dombrowski, Ann-Kathrin and Goel, Shashwat and Li, Nathaniel and Byun, Michael J. and Wang, Zifan and Mallen, Alex and Basart, Steven and Koyejo, Sanmi and Song, Dawn and Fredrikson, Matt and Kolter, J. Zico and Hendrycks, Dan},
  journal = {arXiv:2310.01405},
  year    = {2023}
}

\newpage
\appendix
\newpage
\section{Details of Vulnerability Knowledge Graph Construction}
\label{app:kg_construction}

We construct an enriched vulnerability knowledge graph as the structured evidence source for retrieval and prioritization. The base graph $G_0=(V_0,E_0)$ is built from raw CVE records and contains multiple types of nodes, including CVE, product, vendor, weakness, and reference nodes~\cite{host-etal-2023-constructing}. Each CVE node stores its textual description, severity information such as CVSS scores, affected configurations, weakness categories, and external references. Product and vendor nodes are extracted from affected configurations, weakness nodes represent CWE categories, and reference nodes correspond to advisories, patches, exploit reports, or other supporting evidence. Edges encode relations among these entities, such as which products or vendors a CVE affects, which weakness category it belongs to, and which references support it.

To incorporate prioritization signals beyond raw CVE records, we enrich the base graph with external sources such as known exploited vulnerability lists and public exploit repositories. These sources provide evidence about exploit availability, observed exploitation, patch-related references, and other remediation-relevant signals. The enrichment process updates node attributes and adds edges when new relations are identified:
\begin{equation}
G=\Phi(G_0,S),
\end{equation}
where $S$ denotes the external sources and $\Phi$ is the enrichment procedure. The resulting graph contains both structured vulnerability relations and prioritization-oriented evidence, and is used by the retrieval operator in the main text to form query-specific candidate sets.

\section{Support-Bank Protocol}
\label{app:support_protocol}

Each preference scenario $\gamma\in\{\texttt{importance},\texttt{exploit\_first},\texttt{patch\_order}\}$ has its own support/train/dev/test split over the shared query--candidate instance IDs and its own bank $B_{\gamma}$. The bank stores, for every support instance of scenario $\gamma$, (i)~precomputed normalized score vectors from all view--policy scorers and (ii)~the ranking supervision $y^{(k)}$ for that scenario. Test instance IDs of $\gamma$ never enter $B_{\gamma}$. Main results use the matched bank $B_{\gamma}$ rather than a pooled bank. At inference time, attaching $B_{\gamma}$ provides historical labeled examples for fusion: the scoring prompt has no scenario name, and the model does not predict one. After query-conditioned policy selection, fusion uniformly samples three supports from $B_{\gamma}$ under the selected policy tuple and fits view weights on those supports' labels.

To measure the value of scenario-matched supports, we additionally evaluate a Mixed Support Bank control. We build a pooled bank $B_{\mathrm{mix}}=\bigcup_{\gamma}B_{\gamma}$ and, at test time for every scenario, sample $|S(q)|=3$ supports from $B_{\mathrm{mix}}$ without using the evaluated scenario identity. Adapters, policy selection, $\alpha$, and decoding settings are unchanged. Results are reported in Table~\ref{tab:ablation_results}.

Splits are by query--candidate instance ID within each scenario, with support and test IDs disjoint. %Policy feature-weight templates used to construct adapter training targets are related in emphasis to the evaluation scenarios, but they are not the test annotation guidelines and are never applied to build test rankings.
Policy feature-weight templates are used only to construct adapter training targets. Although both the templates and the evaluation scenarios are motivated by common cybersecurity remediation objectives, the evaluation scenarios define independent human annotation guidelines and are not derived from the policy templates.

\section{Implementation Details}
\label{app:implementation}

We implement HARP with transformer-based causal language models as the backbone. The backbone parameters are frozen, and each policy is implemented as a lightweight adapter on a target MLP projection layer. For example, for Qwen2.5-7B-Instruct, we use \texttt{model.layers.14.mlp.up\_proj} as the target layer. This design allows different policies to learn different scoring preferences while sharing the same frozen backbone LLM.

During training, we optimize only the policy-specific adapters. Each view-policy pair is trained with its own ranking supervision. The adapters are trained for 3 epochs with learning rate $3\times10^{-4}$ using AdamW~\cite{adamw}. The model output is constrained by the \texttt{ranked\_zscore} format described in Appendix~\ref{app:prompt}. The maximum number of new tokens for score generation is set to 200, the decoding temperature is set to 0.75, and invalid score outputs are retried up to 5 times.

The framework uses three prioritization views: \emph{global}, \emph{enterprise}, and \emph{user}, with five, four, and three policies respectively in our benchmark (Table~\ref{tab:global_policy}). These uneven counts reflect the different prioritization trends available under each view rather than an architectural constraint; policies can be added or removed without changing the overall pipeline. Policy selection compares the query embedding with a key derived from each policy's textual name. The fusion weight vector therefore has dimension $3$. For each support instance, we precompute and store the normalized score vectors produced by all policy-specific scorers under all three views. At inference time, after the query selects one policy for each view, fusion samples $|S(q)|=3$ support instances that contain the selected policy configuration and fits the view weights as described in the main text. This fitting step is cheap relative to the per-view LLM scoring calls. The support bank is fixed during evaluation only to keep a clean protocol; it can be refreshed in deployment. Together with uniform multi-view regularization ($\alpha{=}0.3$ in the main experiments) and Top-$K$ consensus routing, this design avoids relying on a single static fusion rule.

For each query instance, candidate-level scores are normalized within the candidate set before fusion. The evaluation cutoff is defined as
\begin{equation}
    K=\lceil 0.3 \cdot |C(q)| \rceil,
\end{equation}
where $C(q)$ is the candidate set for query $q$. We evaluate ranked lists using Precision@$K$, MAP@$K$, and MRR@$K$. Let $G_K(q)$ denote the set of ground-truth Top-$K$ candidates for query $q$, and let $\hat{\pi}_q$ denote the predicted ranking. Precision@$K$ measures the fraction of the predicted Top-$K$ candidates that appear in $G_K(q)$:
\begin{equation}
    \mathrm{Precision@}K(q)
    =
    \frac{|\hat{\pi}_q[1:K]\cap G_K(q)|}{K}.
\end{equation}

MAP@$K$ measures ranking quality within the top-$K$ results by averaging precision at the ranks where a ground-truth Top-$K$ candidate appears:
\begin{equation}
\begin{aligned}
    \mathrm{AP@}K(q)
    &=
    \frac{1}{K}
    \sum_{j=1}^{K}
    \mathrm{Precision@}j(q) \\
    &\quad \cdot
    \mathbb{I}\{\hat{\pi}_q[j]\in G_K(q)\}.
\end{aligned}
\end{equation}
MAP@$K$ is the mean of $\mathrm{AP@}K(q)$ over all test queries.

For MRR@$K$, we treat each candidate in the ground-truth Top-$K$ set as an individual target. For a query $q$, let $r_q(c)$ denote the rank position of candidate $c$ in the predicted ranking $\hat{\pi}_q$. The reciprocal rank of candidate $c$ is
\begin{equation}
    \mathrm{RR}(q,c)=\frac{1}{r_q(c)}.
\end{equation}
We then average the reciprocal ranks over all candidates in the ground-truth Top-$K$ set:
\begin{equation}
    \mathrm{MRR@}K(q)
    =
    \frac{1}{K}
    \sum_{c\in G_K(q)}
    \frac{1}{r_q(c)}.
\end{equation}
The final MRR@$K$ is the mean of $\mathrm{MRR@}K(q)$ over all test queries. If a ground-truth Top-$K$ candidate does not appear in the predicted ranking, its reciprocal rank is set to zero.

We conduct experiments on a Linux cluster using a conda environment. Each experiment is run on NVIDIA A100 GPUs with 80GB memory. Main ranking results for HARP and all baselines are reported as averages over five independent runs. Appendix~\ref{app:run_variance} reports the corresponding standard deviations for HARP across these runs.
For LLM-based baselines, self-consistency uses multiple sampled rankings and aggregates them into the final result. For embedding-based ranking, we use sentence-transformer embeddings by default and rank candidates according to their similarity to the query representation.

Regarding inference cost, evaluating one test query with HARP takes approximately 50--70 seconds on an A100 GPU, depending on the backbone model and candidate-set size. The dominant cost is the LLM scoring calls: HARP invokes the policy-conditioned scorer once per view, so the latency is comparable to multi-call baselines such as LLM-SC and higher than single-call LLM prompting. Other stages are comparatively lightweight or can be amortized: KG retrieval uses indexed graph evidence, support-bank score vectors are precomputed offline, and adaptive fusion only fits a small three-dimensional view-weight vector over sampled support instances.

We evaluate HARP with three backbone LLMs to test whether the proposed framework is robust across different model families and capacities. \textbf{GPT-OSS-20B} is an open-weight model from OpenAI with 21B total parameters and 3.6B active parameters. We include it as a stronger medium-scale backbone model. \textbf{Llama3-8B} is an instruction-tuned model from the Llama 3 family~\cite{llama3herd}. We use it as a representative general-purpose open LLM at the 8B scale. \textbf{Qwen2.5-7B-Instruct} is an instruction-tuned model from the Qwen2.5 series~\cite{qwen25}. We include it as a strong 7B-scale backbone with strong instruction-following capabilities. Across all three models, the backbone parameters are kept frozen, and HARP only trains policy-specific adapters. This setting allows us to compare how the same graph-grounded and preference-adaptive ranking framework behaves under different backbone models.

%Our code is publicly available at \url{https://github.com/HaochenLiu2000/HARP}.
%The datasets used in this paper are available in the public repository.

\section{Dataset Statistics}
\label{app:dataset_scale}

We evaluate on three ranking datasets corresponding to the \emph{importance}, \emph{exploit-first}, and \emph{patch-order} preference scenarios. Each contains $12{,}000$ query--candidate samples with per-CVE ground-truth scores and is split into support ($1{,}200$; 10\%), train ($4{,}800$; 40\%), dev ($1{,}200$; 10\%), and test ($4{,}800$; 40\%). Support instances are used to build the scenario-specific bank $B_{\gamma}$ and to fit fusion weights; train is used for adapter training; and test is used for the reported ranking metrics. All three scenarios share the same query--candidate skeletons and instance-ID split sizes; only the ranking objective used for annotation differs. Candidate-set sizes vary across queries under the natural retrieval distribution, and CVE-level feature summaries are stored separately and indexed by CVE ID.

\section{Run-to-Run Variance of HARP}
\label{app:run_variance}

Table~\ref{tab:run_std} reports the standard deviation of HARP over five independent runs for each backbone and preference scenario. The corresponding mean scores appear in Table~\ref{tab:main_results}. 

\begin{table*}[t]
\centering
\setlength{\tabcolsep}{1.5pt}
\renewcommand{\arraystretch}{1.0}
\caption{Standard deviation of HARP over five independent runs.}
\label{tab:run_std}
\begin{tabular}{lccccccccc}
\toprule
\multirow{2}{*}{\textbf{Backbone}}
& \multicolumn{3}{c}{\textbf{Importance}}
& \multicolumn{3}{c}{\textbf{Exploit\_First}}
& \multicolumn{3}{c}{\textbf{Patch\_Order}} \\
\cmidrule(lr){2-4}\cmidrule(lr){5-7}\cmidrule(lr){8-10}
& \textbf{P@k} & \textbf{MAP@k} & \textbf{MRR@k}
& \textbf{P@k} & \textbf{MAP@k} & \textbf{MRR@k}
& \textbf{P@k} & \textbf{MAP@k} & \textbf{MRR@k} \\
\midrule
GPT-OSS-20B & 0.0082 & 0.0068 & 0.0061 & 0.0084 & 0.0069 & 0.0062 & 0.0079 & 0.0068 & 0.0062 \\
Llama3-8B & 0.0054 & 0.0058 & 0.0056 & 0.0055 & 0.0060 & 0.0057 & 0.0053 & 0.0059 & 0.0056 \\
Qwen2.5-7B-Instruct & 0.0038 & 0.0042 & 0.0040 & 0.0034 & 0.0039 & 0.0039 & 0.0030 & 0.0028 & 0.0036 \\
\bottomrule
\end{tabular}
\end{table*}

\section{Prompt Design and Output Protocol}
\label{app:prompt}

The current prompt style is \texttt{ranked\_zscore}. This format is used to make the LLM output directly usable as a candidate-level ranking signal. The prompt consists of a system instruction that specifies the output contract and a user instruction that provides the query, optional candidate-set synopsis, and candidate CVE list.

The system prompt follows the structure below:
\begin{quote}
You are a CVE prioritization expert. Output ONLY a \texttt{[Scores]} section. Use one line per candidate in the format \texttt{CVE-ID: -0.00}. Scores are normalized within this candidate set and may be signed values. Sort all lines from highest score to lowest score. Include all candidates exactly once.
\end{quote}

The user prompt contains the following fields:
\begin{quote}
\texttt{Question: <query>}\\
\texttt{Candidate-set synopsis: <sample\_summary>} \\
\texttt{CVE candidates:}\\
\texttt{[1] CVE-...}\\
\texttt{[2] CVE-...}\\
\texttt{...}\\
\texttt{Output [Scores] now (sorted highest to lowest):}
\end{quote}

During data generation for policy training, the same prompt format is paired with a target answer in the \texttt{[Scores]} block. The target scores come from the constructed policy-specific ranking supervision, standardized within the candidate set, and sorted from highest to lowest. During inference, the model is required to generate the same \texttt{[Scores]} format. The generated output is parsed by extracting explicit CVE-score pairs. If the output does not contain sufficient valid candidate scores, the system retries generation up to five times. Samples that still fail the output contract are marked invalid.

\nop{\section{Preference Scenarios and Policy Definitions}
\label{app:preference_policy}

This appendix clarifies the difference between \emph{policies} used for adapter training and \emph{preference scenarios} used for testing. A \emph{policy} defines a view-specific scoring preference; the corresponding ranking targets are used only as supervision for adapter training. %A \emph{preference scenario} defines a prioritization objective under which test rankings are manually annotated. 
A \emph{preference scenario} defines a prioritization objective under which ground-truth rankings are annotated by cybersecurity practitioners. At test time, the scoring prompt does not include a preference-scenario ID; fusion instead uses the support bank $B_{\gamma}$ attached to the evaluation run (Appendix~\ref{app:support_protocol}).}

\section{Preference Scenarios and Policy Definitions}
\label{app:preference_policy}

This appendix clarifies the difference between \emph{policies} used for adapter training and \emph{preference scenarios} used for testing. A \emph{policy} defines a view-specific scoring preference; the corresponding ranking targets are used only as supervision for adapter training. A \emph{preference scenario} defines a prioritization objective under which ground-truth rankings are annotated by cybersecurity practitioners. The preference scenarios and their annotation guidelines were developed independently of the policy templates used for adapter training and do not instruct annotators to reproduce any policy weighting scheme. Instead, they define qualitative prioritization objectives based on the available vulnerability evidence rather than predefined feature weights, scoring formulas, or policy-specific ranking rules. At test time, the scoring prompt does not include a preference-scenario ID; fusion instead uses the support bank $B_{\gamma}$ attached to the evaluation run (Appendix~\ref{app:support_protocol}).

We train the scoring modules with policy-specific supervision rather than directly using support-bank labels because the policies capture relatively stable and reusable prioritization behaviors, whereas preference scenarios may vary across organizations and over time. This separation makes HARP a general framework: trained policy adapters can be reused across scenarios, while adaptation to a new or changing scenario requires only updating the support bank and refitting the fusion weights, rather than retraining the scoring modules. Although the policy set can also be extended when needed, it is expected to change less frequently than deployment-specific preference scenarios, making the overall adaptation process more efficient.

\subsection{Candidate Evidence}

For candidate scoring, the LLM scorer observes the natural-language query, the candidate set, the candidate-set synopsis, and candidate-level vulnerability evidence (CVE records, CVSS metadata, exploit signals, references, and product information). Separately, the fusion module observes a small sample of labeled supports from $B_{\gamma}$. The preference scenario is therefore not hidden from the full system: it enters through the support bank used for weight fitting, while remaining absent from the scorer prompt as an explicit tag.

The candidate-level evidence includes textual description, CVSS severity label, CVSS base score, attack vector, exploit status, exploit flag, CWE identifiers, affected product keys, platform information, patch signals from references, reference count, version-related signals, and publication recency. These fields are summarized into the candidate feature summary provided to the LLM. For policy-adapter training, they are also converted into numerical feature contributions that guide the construction of policy-specific ranking supervision.

\subsection{Feature Contribution Functions}

We first define a common set of feature contribution functions $\{\phi_f(c)\}$ for each candidate CVE $c$. These functions map raw vulnerability evidence to normalized numerical values and are used only to guide the construction of policy-specific training targets. Table~\ref{tab:feature_contrib} lists the feature contribution rules used in our implementation.

\begin{table*}[h]
\setlength{\tabcolsep}{2.0pt}
\renewcommand{\arraystretch}{1.0}
\caption{Feature contribution rules used to guide the construction of policy-specific training supervision.}
\label{tab:feature_contrib}
\begin{tabular}{ll}
\toprule
\textbf{Feature} & \textbf{Contribution Rule} \\
\midrule
\texttt{exploit} & $1.0$ if \texttt{KNOWN\_EXPLOITED}; $0.6$ if \texttt{has\_exploit} is true; otherwise $0.0$ \\
\texttt{severity} & If \texttt{base\_score} exists, $\min(1,\texttt{base\_score}/10)$; otherwise use severity mapping \\
\texttt{patch} & $1.0$ if \texttt{has\_patch\_in\_references} is true; otherwise $0.0$ \\
\texttt{attack\_vector} & NETWORK: $1.0$; ADJACENT: $0.5$; LOCAL: $0.2$; PHYSICAL: $0.1$; otherwise $0.0$ \\
\texttt{recency} & $\max(0, 1-\texttt{days\_since\_published}/365)$ \\
\texttt{version\_info} & $0.5 \cdot \texttt{has\_version\_info} + 0.5 \cdot \texttt{affects\_legacy\_versions}$ \\
\texttt{reference} & $\min(1,\texttt{reference\_count}/10)$ \\
\texttt{cwe} & $\min(1,\texttt{len(cwe\_ids)}/5)$ \\
\texttt{platform} & $1.0$ if platform list is non-empty; otherwise $0.0$ \\
\texttt{scope} & $\min(1,\texttt{len(product\_keys)}/10)$ \\
\bottomrule
\end{tabular}
\end{table*}

\begin{table*}[t]
\setlength{\tabcolsep}{1pt}
\renewcommand{\arraystretch}{1.0}
\caption{Policy weights of the three views \emph{global}, \emph{enterprise} and \emph{user}.}
\label{tab:global_policy}
\begin{tabular}{lcccccccccc}
\toprule
\textbf{Policy} & \textbf{exploit} & \textbf{severity} & \textbf{patch} & \textbf{attack} & \textbf{recency} & \textbf{version} & \textbf{reference} & \textbf{cwe} & \textbf{platform} & \textbf{scope} \\
\midrule
\texttt{global\_balanced} & 2.5 & 2.0 & 0.5 & 0.8 & 0.5 & 0.3 & 0.3 & 0.2 & 0.1 & 0.2 \\
\texttt{global\_exploit\_first} & 3.5 & 1.0 & 0.3 & 0.5 & 0.6 & 0.2 & 0.2 & 0.1 & 0.0 & 0.2 \\
\texttt{global\_severity\_first} & 2.0 & 2.5 & 1.0 & 0.5 & 0.3 & 0.5 & 0.2 & 0.2 & 0.1 & 0.2 \\
\texttt{global\_recent} & 1.5 & 1.0 & 0.5 & 0.3 & 2.5 & 0.3 & 0.5 & 0.2 & 0.1 & 0.2 \\
\texttt{global\_version\_aware} & 1.5 & 1.5 & 1.0 & 0.4 & 0.3 & 2.0 & 0.3 & 0.2 & 0.2 & 0.2 \\
\midrule
\texttt{enterprise\_patch} & 2.0 & 1.0 & 2.0 & 0.3 & 0.2 & 0.6 & 0.4 & 0.1 & 0.1 & 0.2 \\
\texttt{enterprise\_compliance} & 2.0 & 2.5 & 0.8 & 0.5 & 0.3 & 0.4 & 0.2 & 0.2 & 0.1 & 0.5 \\
\texttt{enterprise\_scope} & 2.5 & 2.0 & 0.8 & 0.5 & 0.4 & 0.4 & 0.3 & 0.2 & 0.2 & 1.2 \\
\texttt{enterprise\_reference} & 1.5 & 1.2 & 1.0 & 0.4 & 0.5 & 0.4 & 1.5 & 0.2 & 0.1 & 0.2 \\
\midrule
\texttt{user\_risk} & 2.5 & 2.0 & 0.5 & 0.8 & 0.5 & 0.3 & 0.3 & 0.2 & 0.1 & 0.2 \\
\texttt{user\_quick\_fix} & 2.0 & 1.0 & 2.0 & 0.3 & 0.2 & 0.6 & 0.4 & 0.1 & 0.1 & 0.2 \\
\texttt{user\_awareness} & 2.0 & 1.5 & 0.5 & 0.5 & 0.8 & 0.4 & 1.2 & 1.0 & 0.2 & 0.2 \\
\bottomrule
\end{tabular}
\end{table*}

For textual severity labels, we use the following mapping:
\begin{equation}
\begin{aligned}
\mathrm{CRITICAL} &= 1.0, \quad \mathrm{HIGH} = 0.7, \\
\mathrm{MEDIUM} &= 0.3, \quad \mathrm{LOW} = 0.1.
\end{aligned}
\end{equation}
All other severity labels are mapped to $0.0$.

Given any feature-weight vector $\lambda$, we first clip negative feature weights and normalize the remaining positive weights:
\begin{equation}
    \tilde{\lambda}_f=\max(\lambda_f,0),
    \qquad
    \bar{\lambda}_f=
    \frac{\tilde{\lambda}_f}{\sum_{f'}\tilde{\lambda}_{f'}}.
\end{equation}
We then compute a generic weighted feature score:
\begin{equation}
    s_{\lambda}(c)=\max\left(0,\sum_f \bar{\lambda}_f\phi_f(c)\right).
\end{equation}
This equation is a generic scoring template used only when constructing policy-specific training targets.

\subsection{Policy Training Targets}

A policy is a discrete scoring preference template used to train a policy-specific adapter. Policies are organized under the three prioritization views: \emph{global}, \emph{enterprise}, and \emph{user}. Each policy $\pi_{l,r}$ in view $l$ is associated with a feature-weight template $\lambda_{l,r}$ that guides the construction of its ranking supervision. The policy-specific training score for candidate $c$ is defined as
\begin{equation}
    y^{\mathrm{pol}}_{l,r}(c)=s_{\lambda_{l,r}}(c).
\end{equation}
The ranking induced by $y^{\mathrm{pol}}_{l,r}(c)$ is used only to train the adapter corresponding to policy $\pi_{l,r}$. These scores are not the evaluation labels used in the final benchmark, and the policy templates are never applied to construct test rankings.

Different policies emphasize different remediation criteria over the same candidate evidence. For example, an exploit-focused policy assigns larger weights to exploit-related features, while a patch-focused policy assigns larger weights to patch-related features. The feature-weight templates used to guide construction of all policy adapters are listed in Table~\ref{tab:global_policy}.

\subsection{Scenario-Based Test Annotation}
\label{app:scenario_labels}

%Preference scenarios define the evaluation objectives: \emph{importance}, \emph{exploit\_first}, and \emph{patch\_order}. For each scenario, we manually annotate rankings from the query, candidate evidence, and a short scenario guideline; policy feature-weight templates used for adapter training are not used during this annotation. The same query--candidate instances can therefore receive different rankings under different scenarios. All annotated rankings are re-checked multiple times, and remaining disagreements are resolved to a single consensus ranking. These test rankings are independent of policy-training templates: the templates guide adapter supervision only, are not used to construct test rankings, and are related in emphasis to the evaluation scenarios but not identical to them. At scoring time the prompt has no scenario ID; fusion uses $B_{\gamma}$ as described in Appendix~\ref{app:support_protocol}.

Preference scenarios define the evaluation objectives: \emph{importance}, \emph{exploit\_first}, and \emph{patch\_order}. For each scenario, ground-truth rankings were annotated by cybersecurity practitioners from our organization's IT/security team using the query, candidate evidence, and a scenario-specific prioritization guideline. The same query--candidate instances can therefore receive different rankings under different scenarios. Annotators independently assessed the candidate CVEs according to their professional judgment, and remaining disagreements were resolved through discussion to obtain a consensus ranking. The policy feature-weight templates used for adapter training were not available to the annotators and were never used to generate evaluation labels. Instead, these templates served solely to construct policy-specific supervision for training the adapters. %These test rankings are independent of policy-training templates: the templates guide adapter supervision only, are not used to construct test rankings, and are related in emphasis to the evaluation scenarios but not identical to them. At scoring time, the prompt has no scenario ID; fusion uses $B_{\gamma}$ as described in Appendix~\ref{app:support_protocol}.
These test rankings are independent of the policy-training templates. The templates provide supervision only for adapter training, whereas the evaluation scenarios define qualitative human annotation guidelines. Although both are motivated by common cybersecurity remediation objectives, the evaluation scenarios were designed independently and are not derived from the policy templates.

\section{Overall Algorithm}
\label{app:algorithm}

Algorithm~\ref{alg:overall} summarizes the HARP inference procedure.

\begin{algorithm}[t]
\caption{HARP: Hierarchical Adaptive Ranking with Preference-Adaptive Fusion for Query-Based CVE Prioritization}
\label{alg:overall}
\begin{algorithmic}[1]
\REQUIRE Query $q$, graph $G$, policy sets $\{\Pi_l\}$, adapters $\mathcal{A}$, preference-scenario support bank $B_{\gamma}$, coefficient $\alpha$
\ENSURE Ranked CVE list $\hat{\pi}_q$

\STATE $C \leftarrow R(q,G)$ \hfill \COMMENT{Retrieve candidate CVEs}
\STATE $x \leftarrow \mathrm{EvidenceBuilder}(q,C)$ \hfill \COMMENT{Build candidate feature summary}

\FOR{$l=1,\ldots,L$}
    \STATE $\pi_l^* \leftarrow \mathrm{PolicySelect}(q,\Pi_l)$
    \STATE $a_l \leftarrow \mathrm{Score}(q,C,x,\pi_l^*;\mathcal{A})$
    \STATE $z_l \leftarrow \mathrm{Normalize}(a_l)$
\ENDFOR

\STATE $\boldsymbol{\pi}^* \leftarrow (\pi_1^*,\ldots,\pi_L^*)$
\STATE $\mathcal{S} \leftarrow \mathrm{SampleSupport}(B_{\gamma},\boldsymbol{\pi}^*)$
\STATE $w \leftarrow \mathrm{FitFusionWeights}(\mathcal{S})$

\STATE $F_{\mathrm{adapt}} \leftarrow \mathrm{WeightedFuse}(\{z_l\}_{l=1}^{L},w)$
\STATE $F_{\mathrm{uni}} \leftarrow \mathrm{UniformFuse}(\{z_l\}_{l=1}^{L})$
\STATE $F \leftarrow \alpha F_{\mathrm{adapt}}+(1-\alpha)F_{\mathrm{uni}}$

\IF{$\mathrm{TopKAgree}(\{z_l\}_{l=1}^{L})$}
    \STATE $\hat{\pi}_q \leftarrow \mathrm{ConsensusTopK}(\{z_l\}_{l=1}^{L},F)$
\ELSE
    \STATE $\hat{\pi}_q \leftarrow \mathrm{RankByScore}(F)$
\ENDIF

\RETURN $\hat{\pi}_q$
\end{algorithmic}
\end{algorithm}

\section{Sensitivity Analysis on $\alpha$}
\label{app:alpha_sensitivity}

\begin{figure*}[t]
\centering
\includegraphics[width=1\linewidth]{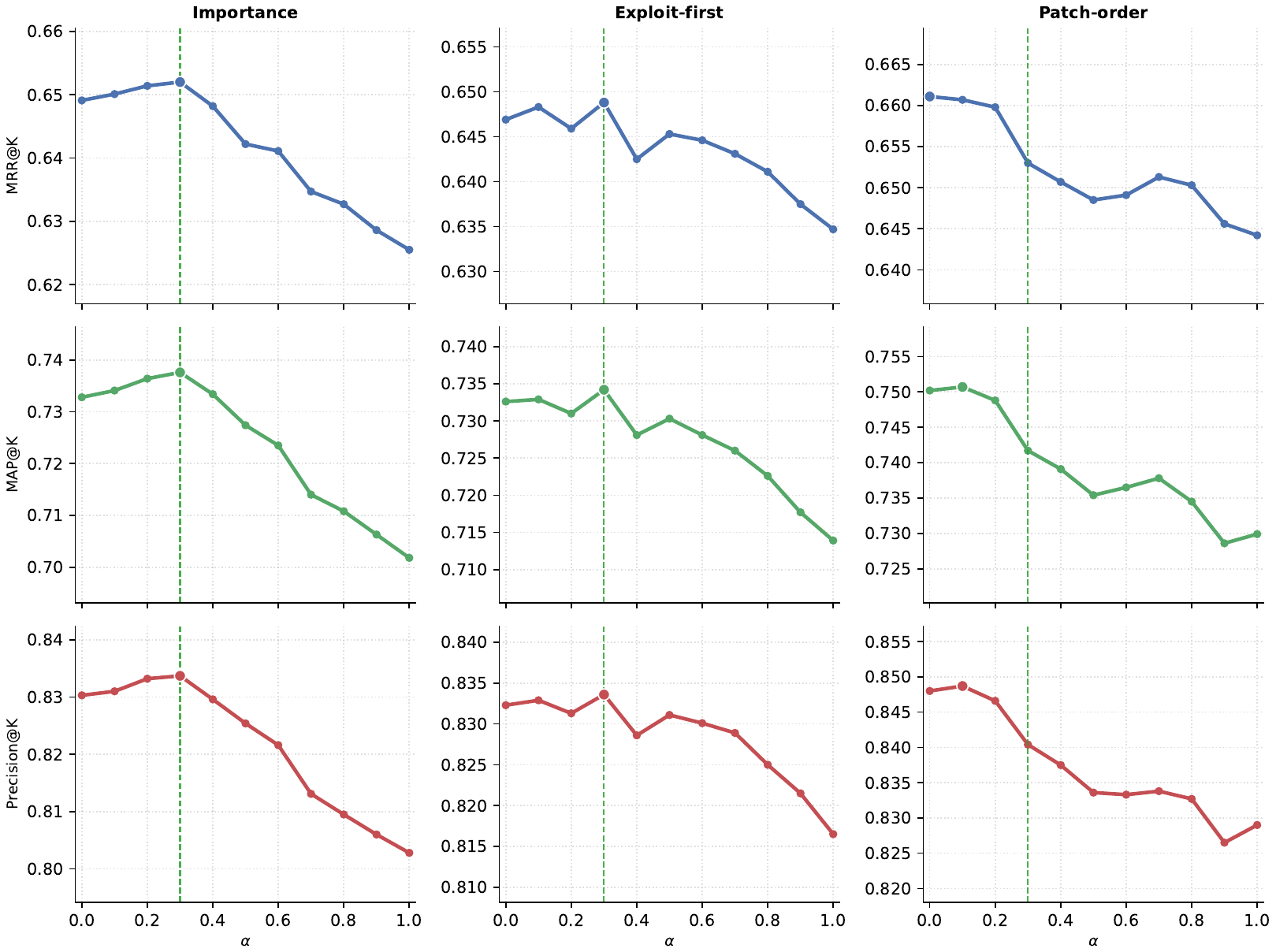}
    \caption{Sensitivity analysis of the interpolation coefficient $\alpha$ on Qwen2.5-7B-Instruct. We report Precision@K, MAP@K, and MRR@K under the importance, exploit-first, and patch-order preference scenarios. The dashed vertical line indicates the value used in the main experiments, $\alpha=0.3$.}
    \label{fig:alpha}
\end{figure*}

We further study the effect of the interpolation coefficient $\alpha$ in the final score
$F(q,c_i)=\alpha F_{\mathrm{adapt}}(q,c_i)+(1-\alpha)F_{\mathrm{uni}}(q,c_i)$.
Figure~\ref{fig:alpha} reports the sensitivity results on Qwen2.5-7B-Instruct across the three preference scenarios. The dashed vertical line marks the value used in our main experiments, $\alpha=0.3$.

The effect of $\alpha$ differs across preference scenarios. In the \emph{importance} preference scenario, all three metrics increase slightly when $\alpha$ grows from $0$ to $0.3$, and then decrease as $\alpha$ becomes larger. This suggests that a moderate amount of adaptive fusion improves overall-priority ranking, while relying too heavily on adaptive fusion weakens the stabilizing effect of the uniform multi-view term. In the \emph{exploit-first} preference scenario, the curves are more non-monotonic, but the best or near-best results still appear around $\alpha=0.3$ for MRR@K and Precision@K, with MAP@K remaining competitive. This indicates that exploit-focused ranking benefits from adaptive view weighting, but the improvement is not strictly monotonic.

The \emph{patch-order} preference scenario shows a different trend. Performance is strongest when $\alpha$ is small and generally decreases as $\alpha$ increases. This suggests that patch-oriented ranking benefits more from preserving balanced evidence across views, and excessive adaptive weighting may overemphasize one view at the cost of other useful patch-related signals. Overall, $\alpha$ trades off support-conditioned adaptive fusion against uniform multi-view stability. We select $\alpha{=}0.3$ on the development split as a robust operating point across preference scenarios; the optimum can still vary by objective (e.g., patch-order favors smaller $\alpha$).

% Required packages:
% \usepackage{tcolorbox}
% \tcbuselibrary{skins,breakable}
% \usepackage{enumitem,xcolor,amsmath,amssymb}

% ============================================================
% Case-study box definitions
% ============================================================

\newcommand{\fusionwidebox}[2][]{%
  \begin{figure*}[t]
  \begin{tcolorbox}[%
    #1,
    enhanced,
    width=\textwidth,
    colback=gray!4,
    colframe=gray!50,
    boxrule=0.4pt,
    arc=1.5mm,
    left=4pt,
    right=4pt,
    top=4pt,
    bottom=4pt,
    fontupper=\small,
    fonttitle=\bfseries
  ]
  #2
  \end{tcolorbox}
  \end{figure*}%
}

\newcommand{\caseinnerbox}[2]{%
  \begin{tcolorbox}[
    title={#1},
    colback=white,
    colframe=gray!45,
    boxrule=0.35pt,
    arc=1mm,
    left=3pt,
    right=3pt,
    top=2pt,
    bottom=2pt,
    fonttitle=\bfseries\footnotesize,
    fontupper=\footnotesize,
    width=\linewidth
  ]
  #2
  \end{tcolorbox}
}

\newcommand{\gtinnerbox}[1]{%
  \begin{tcolorbox}[
    title={Ground Truth},
    colback=green!4,
    colframe=green!45!black,
    boxrule=0.45pt,
    arc=1mm,
    left=3pt,
    right=3pt,
    top=2pt,
    bottom=2pt,
    fonttitle=\bfseries\footnotesize,
    fontupper=\footnotesize,
    width=\linewidth
  ]
  #1
  \end{tcolorbox}
}

\newcommand{\predinnerbox}[1]{%
  \begin{tcolorbox}[
    title={View Predictions and Fusion},
    colback=blue!2,
    colframe=blue!35!black,
    boxrule=0.4pt,
    arc=1mm,
    left=3pt,
    right=3pt,
    top=2pt,
    bottom=2pt,
    fonttitle=\bfseries\footnotesize,
    fontupper=\footnotesize,
    width=\linewidth
  ]
  #1
  \end{tcolorbox}
}

% Prediction table
\newcommand{\predtable}[1]{%
  \predinnerbox{%
    \begin{tabularx}{\linewidth}{
      @{}
      >{\raggedright\arraybackslash}p{0.13\linewidth}
      >{\raggedright\arraybackslash}X
      >{\raggedright\arraybackslash}p{0.25\linewidth}
      @{}
    }
    #1
    \end{tabularx}%
  }%
}

% Analysis box
\newcommand{\analysisbox}[1]{%
  \begin{tcolorbox}[
    title={Analysis},
    colback=yellow!3,
    colframe=yellow!45!black,
    boxrule=0.4pt,
    arc=1mm,
    left=3pt,
    right=3pt,
    top=2pt,
    bottom=2pt,
    fonttitle=\bfseries\footnotesize,
    fontupper=\footnotesize,
    width=\linewidth
  ]
  #1
  \end{tcolorbox}
}

% ============================================================
% Utility commands
% ============================================================

\newcommand{\cve}[1]{\texttt{CVE-#1}}

\newcommand{\good}{%
  \textcolor{green!50!black}{\ensuremath{\checkmark}}%
}

\newcommand{\bad}{%
  \textcolor{red!70!black}{\ensuremath{\times}}%
}

\newcommand{\gtmark}{%
  \textcolor{green!50!black}{\ensuremath{\checkmark}}%
}

\newcommand{\ra}{%
  \ensuremath{\rightarrow}%
}

% ============================================================
% Case Studies
% ============================================================

\section{Case Studies}
\label{app:case_study}

We present two cases under Qwen2.5-7B-Instruct with $\alpha=0.3$
to examine how HARP combines view-specific rankings.
Each case reports the ranking question, a brief candidate-set synopsis,
the ground-truth ranking, and the rankings produced by the global,
enterprise, user, and fusion scores.
We focus on top-$K$ behavior: which candidate each view places first,
how the fused ranking differs from the single-view rankings,
and what the fitted view weights imply about relative view contributions
under the selected policy configuration.

The case boxes below follow the same structure.
A check mark indicates that the corresponding view or fusion result
ranks the ground-truth top candidate at rank 1,
while a cross indicates otherwise.

% ============================================================
% Case I
% ============================================================

\fusionwidebox[
  title={Case I --- \texttt{cve\_rank\_010363} ($n=10$)}
]{

\caseinnerbox{Ranking Question}{
Which vulnerabilities associated with Goodlayers should be addressed
first to minimize risk?
}

\caseinnerbox{Candidate-Set Synopsis}{
The candidate set contains legacy Travelon Express XSS/upload issues
and recent GoodLayers vulnerabilities with vendor fixes.
The patch-order ground truth favors recent patched GoodLayers CVEs,
especially \cve{2025-39503} and \cve{2025-59580}.
}

\gtinnerbox{
\textbf{Full ranking:}
\cve{2025-39503}\gtmark\ (0.559)
\ra \cve{2025-59580} (0.555)
\ra \cve{2025-39502} (0.497)
\ra \cve{2025-53342} (0.494)
\ra \cve{2008-5864} (0.406)
\ra \cve{2024-9458} (0.406)
\ra \cve{2008-0184} (0.403)
\ra \cve{2009-4617} (0.403)
\ra \cve{2012-2938},
\cve{2012-2939} (0.230).
}

\predtable{
Global &
\cve{2008-0184}
\ra \cve{2009-4617}
\ra \cve{2008-5864} &
GT\#1 rank: 8; Top-1: \bad
\\

Enterprise &
\cve{2008-0184}
\ra \cve{2009-4617}
\ra \cve{2025-39503} &
GT\#1 rank: 3; Top-1: \bad
\\

User &
\cve{2025-59580}
\ra \cve{2025-39503}
\ra \cve{2025-53342} &
GT\#1 rank: 2; Top-1: \bad
\\

Fusion &
\textbf{\cve{2025-39503}}
\ra \textbf{\cve{2025-59580}}
\ra \textbf{\cve{2025-39502}} &
GT\#1 rank: 1; Top-1: \good
\\
}

\analysisbox{
In this case, only fusion places GT\#1 first and also matches
the GT top-3 order.
The global view ranks legacy CVEs highly, whereas enterprise and user
place more weight on recent patched GoodLayers CVEs.
The fitted weights
$(w_g,w_e,w_u)=(-0.16,+0.36,+0.48)$
reflect this pattern by down-weighting the global view
and increasing the contributions of enterprise and user.
}

}

% ============================================================
% Case II
% ============================================================

\fusionwidebox[
  title={Case II --- \texttt{cve\_rank\_011853} ($n=10$)}
]{

\caseinnerbox{Ranking Question}{
Which vulnerabilities from GE should be prioritized based on critical
impact or widespread usage of their products?
}

\caseinnerbox{Candidate-Set Synopsis}{
The candidate set contains industrial GE vulnerabilities and older web CVEs.
The patch-order ground truth favors \cve{2014-0750},
a path traversal vulnerability with exploit evidence,
over several high-CVSS but lower-ranked industrial entries.
}

\gtinnerbox{
\textbf{Full ranking:}
\cve{2014-0750}\gtmark\ (0.461)
\ra \cve{2011-1565} (0.415)
\ra \cve{2006-1216} (0.412)
\ra \cve{2015-8735} (0.412)
\ra \cve{2006-6198} (0.409)
\ra \cve{2009-3967} (0.403)
\ra \cve{2020-36549} (0.321)
\ra \cve{2022-2948} (0.265)
\ra \cve{2017-17576} (0.218)
\ra \cve{2018-10613} (0.049).
}

\predtable{
Global &
\cve{2006-6198} \ra \textit{...} &
GT\#1 rank: 6; Top-1: \bad
\\

Enterprise &
\cve{2006-1216} \ra \textit{...} &
GT\#1 rank: 5; Top-1: \bad
\\

User &
\cve{2011-1565}
\ra \cve{2014-0750}
\ra \textit{...} &
GT\#1 rank: 2; Top-1: \bad
\\

Fusion &
\textbf{\cve{2014-0750}}
\ra \textbf{\cve{2011-1565}}
\ra \textbf{\cve{2015-8735}} &
GT\#1 rank: 1; Top-1: \good
\\
}

\analysisbox{
Here, none of the single views ranks GT\#1 first,
but fusion places \cve{2014-0750} at the top.
The fused ranking therefore does not reduce to majority voting
over view-wise top-1 predictions;
instead, support-based weighting recombines partial evidence across views
under the selected policy configuration.
}

}

\section{Use of AI Assistance}
\label{sec:ai_assistance}

We used ChatGPT only as a writing assistance tool to improve the clarity, grammar, and readability of the manuscript. The tool was not used to generate research ideas, design the method, conduct experiments, produce experimental results, or draw scientific conclusions. All technical content, experimental design, implementation, analysis, and final claims were created and verified by the authors.

\end{document}